\documentclass[EPiC]{easychair}

\usepackage{doc}
\usepackage{lipsum}
\usepackage{holtexbasic,proof}
\usepackage{amsmath}
\usepackage{stmaryrd}
\usepackage[utf8]{inputenc}
\usepackage[T1]{fontenc}
\usepackage{mathpartir}
\usepackage{microtype}

\newcommand{\dependencypair}[3]{\ensuremath{#2\;\mathrel{\rightsquigarrow_{#1}}\;#3}}

\newcommand{\HOLTokenSupCirc}{\ensuremath{{}^\circ}}
\newcommand{\HOLTokenSupBullet}{\ensuremath{{}^\bullet}}

\renewcommand{\HOLTokenTurnstile}{\ensuremath{\vdash\!\!}}
\renewcommand{\HOLConst}[1]{\textsf{\small #1}}
\newcommand{\HOLConstSub}[1]{\textsf{#1}}

\renewcommand{\HOLSymConst}[1]{\HOLConst{#1}}
\renewcommand{\HOLTyOp}[1]{\HOLConst{#1}}
\renewcommand{\HOLinline}[1]{\textsf{\ensuremath{#1}}}
\renewcommand{\HOLKeyword}[1]{\mathsf{#1}}
\renewcommand{\HOLTokenBar}{\ensuremath{\mathtt{|}}}

\usepackage{environ}
\NewEnviron{holthmenv}{%
  \[
  \scalebox{1.0}{\begin{array}[t]{l}
  \BODY
  \end{array}}
  \]}

\NewEnviron{holthmenvl}{%
\begin{equation}\begin{aligned}
  \scalebox{0.9}{\begin{array}[t]{l}\BODY\end{array}}
\end{aligned}\end{equation}}

\title{A Mechanised Semantics for HOL with Ad-hoc Overloading}

\author{
Johannes {\AA}man Pohjola\inst{1}\inst{2}
\and
Arve Gengelbach\inst{3}
}

\institute{
  CSIRO's Data61,
  Sydney, Australia\\
  \email{johannes.amanpohjola@data61.csiro.au}
\and
  University of New South Wales,
  Sydney, Australia
\and
   Uppsala University,
   Uppsala, Sweden\\
   \email{arve.gengelbach@it.uu.se}
 }

\authorrunning{{\AA}man Pohjola and Gengelbach}

\titlerunning{A Mechanised Semantics for HOL with Ad-hoc Overloading}

\begin{document}

\maketitle

\begin{abstract}
  Isabelle/HOL augments classical higher-order logic with ad-hoc overloading of
  constant definitions---
  that is, one constant may have several definitions for non-overlapping types.
  In this paper, we present a mechanised proof
  that HOL with ad-hoc overloading is consistent.
  All our results have been formalised in the HOL4 theorem prover.
\end{abstract}


%
%

\section{Introduction}
\label{sec:introduction}

The consistency of higher-order logic (HOL), and the correctness of its
implementations, is the bedrock supporting the truth claims of landmark success stories in formal verification such as seL4~\cite{DBLP:conf/sosp/KleinEHACDEEKNSTW09}, Flyspeck~\cite{flyspeck} and CakeML~\cite{DBLP:conf/popl/KumarMNO14}. And the bedrock is firm: there is something of a tradition within the HOL community, from Harrison~\cite{DBLP:conf/cade/Harrison06} to Kumar et al.~\cite{DBLP:conf/itp/KumarAMO14,DBLP:journals/jar/KumarAMO16}, of eating our own dog food by using HOL provers to verify the meta-theory of HOL itself. Such self-verification efforts must always come with an asterisk, lest we run afoul of self-reference paradoxes~\cite{godel}; nonetheless, the trust story is significantly strengthened.

At the time of writing, Isabelle/HOL~\cite{nipkow:isabelle} is by far the most popular HOL implementation: at the most recent ITP conference~\cite{DBLP:conf/itp/2019}, traditionally the main venue for HOL theorem proving, Isabelle/HOL papers outnumbered papers about other HOL provers 13 to 3.\footnote{One paper described the emulation of HOL4 in Isabelle/HOL~\cite{DBLP:conf/itp/ImmlerRW19}. We count this in both columns.} It is a shame, then, that the above self-verification efforts do not quite apply to Isabelle/HOL, because Isabelle has more expressive primitives for introducing definitions. In particular, Isabelle, unlike the other HOLs, admits \emph{ad-hoc overloading} of polymorphic constants. This significantly complicates the trust story: the meaning of previously introduced constants may be changed by future overloads, so extra care must be taken to avoid overloads that introduce circularities or contradictions. In this paper, we strengthen the bedrock for Isabelle/HOL by presenting the first machine-checked proof of consistency for higher-order logic with ad-hoc overloading.

This continues and unifies two previous lines of work.
Ad-hoc overloading in the context of Isabelle/HOL has been studied by Wenzel~\cite{DBLP:conf/tphol/Wenzel97}, by Obua~\cite{DBLP:conf/rta/Obua06}, and most recently by Kun{\v{c}}ar and Popescu~\cite{kuncar2019}. Our consistency proof is based on the pen-and-paper consistency proof by Kun{\v{c}}ar and Popescu. Our formalisation, meanwhile, uses the HOL4 formalisation of HOL with definitions (nicknamed Candle) by Kumar et al.~\cite{DBLP:conf/itp/KumarAMO14} as its starting point.

But our work is not a straightforward mechanisation of the former, nor is it a straightforward extension of the latter. From Candle we inherit most of the syntax and some generic infrastructure, but because the model construction of Kun{\v{c}}ar and Popescu differs so much from the standard Pitts-style construction that Candle follows, we start essentially from scratch on the semantics. We find two (fixable) mistakes in Kun{\v{c}}ar and Popescu's proof regarding the treatment of variable names and substitution, forcing us to innovate by e.g.~formulating a novel style of term semantics using lazy evaluation of ground type substitutions. Furthermore, we consider signature extensions in addition to definitional extensions, and use less minimalist definitional mechanisms: constant definitions in the style of Arthan~\cite{DBLP:conf/itp/Arthan14}, and type definitions requiring proof that the new type is inhabited up-front. These introduce new complications by making the model construction depend on soundness.

The impact of this work is as follows. First, we strengthen the trust story for the Isabelle/HOL logic by formally validating (and, on rare occasions, fixing) the novel and original model construction by Kun{\v{c}}ar and Popescu. Second, since our logic is a conservative extension of Candle, we show that the Candle consistency proof is robust to a different style of model construction. Finally, our contributions lay the groundwork for extracting a future verified CakeML implementation of an Isabelle/HOL kernel.



We delimit our scope as follows. Unlike Kumar et al., we do not currently leverage our proofs by extracting a verified CakeML implementation of our kernel. Doing so requires verifying Kun{\v{c}}ar's algorithm for checking orthogonality and termination of dependency relations~\cite{DBLP:conf/cpp/Kuncar15}; this will be the topic of future work.
Unlike Kun{\v{c}}ar and Popescu, we do not consider ad-hoc overloading of type constructors---while an intriguing concept, this is a feature that Isabelle/HOL does not have, nor are we aware of a use case that would justify its inclusion.
Finally, we alluded earlier to an asterisk to avoid self-reference paradoxes. The asterisk is that our consistency proof for HOL with the axiom of infinity is predicated upon assumptions, stating the existence of a set theory, that cannot be witnessed in HOL4 without additional axioms.

All definitions and theorems in this paper are formalised in the HOL4 theorem prover, and the proof scripts are available online.\footnote{\url{https://code.cakeml.org/tree/master/candle/overloading}}

\section{Syntax}
\label{sec:syntax}

In this section, we present the syntax of higher-order logic that we base our work on.
Much of it is inherited from Kumar et al.~\cite{DBLP:conf/itp/KumarAMO14}, who in turn inherited
much from Harrison~\cite{DBLP:conf/cade/Harrison06}; we will explicitly point out the places where we differ.
But first, a few words on notation.

\subsection{Notation}

For the most part, mathematical formulas are generated from the HOL4 formalisation.
Constant symbols are written in \HOLConst{this font}, variables in \HOLinline{\HOLFreeVar{this}\;\HOLFreeVar{font}},
and string literals are \HOLinline{\HOLStringLitDG{bracketed like this}}. Colon means ``type of'',
as in $\HOLinline{\HOLConst{map}}:\HOLinline{(\ensuremath{\alpha}\;\ensuremath{\Rightarrow}\;\ensuremath{\beta})\;\ensuremath{\Rightarrow}\;\ensuremath{\alpha}\;\HOLTyOp{list}\;\ensuremath{\Rightarrow}\;\ensuremath{\beta}\;\HOLTyOp{list}}$.
Function update \HOLinline{\HOLFreeVar{f}\ensuremath{\llparenthesis}\HOLFreeVar{x}\;\mapsto\;\HOLFreeVar{y}\ensuremath{\rrparenthesis}} denotes the function that is like \HOLinline{\HOLFreeVar{f}} except it maps
\HOLinline{\HOLFreeVar{x}} to \HOLinline{\HOLFreeVar{y}}. \HOLinline{\HOLSymConst{++}} denotes list append.
We will silently convert lists to sets when convenient.
The sum type \HOLinline{\ensuremath{\alpha}\;\ensuremath{+}\;\ensuremath{\beta}} is the disjoint union of \HOLinline{\ensuremath{\alpha}} and \HOLinline{\ensuremath{\beta}}, whose
elements are tagged \HOLinline{\HOLConst{INL}} and \HOLinline{\HOLConst{INR}}, respectively.
To avoid confusion between HOL4 notation and the deep embedding of HOL syntax within HOL4,
the latter will be presented in AST form.
Theorems proved in HOL4 are marked with an initial $\vdash$ before the theorem statement;
when $\vdash$ is used as an infix symbol, it instead denotes provability in the (deeply
embedded) HOL inference system.

\subsection{Types and terms}

The types and terms are those of the $\lambda$-calculus with rank 1 polymorphism:
\begin{holthmenv}
\HOLTyOp{type}\;=\;\HOLConst{Tyvar}\;\HOLTyOp{string}\;\HOLTokenBar{}\;\HOLConst{Tyapp}\;\HOLTyOp{string}\;(\HOLTyOp{type}\;\HOLTyOp{list})\\[0.5em]
\HOLTyOp{term}\;=\;\HOLConst{Var}\;\HOLTyOp{string}\;\HOLTyOp{type}\;\HOLTokenBar{}\;\HOLConst{Const}\;\HOLTyOp{string}\;\HOLTyOp{type}\;\HOLTokenBar{}\;\HOLConst{Comb}\;\HOLTyOp{term}\;\HOLTyOp{term}\;\HOLTokenBar{}\;\HOLConst{Abs}\;\HOLTyOp{term}\;\HOLTyOp{term}
\end{holthmenv}
\noindent
In the $\lambda$-abstraction \HOLinline{\HOLConst{Abs}\;\ensuremath{\HOLFreeVar{t}\sb{\mathrm{1}}}\;\ensuremath{\HOLFreeVar{t}\sb{\mathrm{2}}}}, the binder \HOLinline{\ensuremath{\HOLFreeVar{t}\sb{\mathrm{1}}}} is a term for uniformity.
In practice we only consider well-formed terms,
where \HOLinline{\ensuremath{\HOLFreeVar{t}\sb{\mathrm{1}}}\;\HOLSymConst{=}\;\HOLConst{Var}\;\HOLFreeVar{x}\;\HOLFreeVar{ty}} for some $\HOLinline{\HOLFreeVar{x}},\HOLinline{\HOLFreeVar{ty}}$.
Note that names are just strings. We do not use any binding framework,
such as nominal logic~\cite{PittsAM:newaas-jv} or higher-order abstract syntax~\cite{DBLP:conf/pldi/PfenningE88}.
This is to facilitate future code extraction to CakeML.
The treatment of $\alpha$-conversion will not be relevant in this paper;
the interested reader may consult Kumar et al.~\cite{DBLP:journals/jar/KumarAMO16}.

The built-in types and constants (booleans, functions and equality) are definable by the following abbreviations:

\[\small
\begin{array}{l@{\quad}r@{\quad}l}
  \HOLinline{\HOLConst{Bool}}&  \mbox{for} & \HOLinline{\HOLConst{Tyapp}\;\HOLStringLitDG{bool}\;[]}\\
  \HOLinline{\HOLConst{Fun}\;\HOLFreeVar{x}\;\HOLFreeVar{y}} & \mbox{for} & \HOLinline{\HOLConst{Tyapp}\;\HOLStringLitDG{fun}\;[\HOLFreeVar{x};\;\HOLFreeVar{y}]}\\
  \HOLinline{\HOLConst{Equal}\;\HOLFreeVar{ty}} & \mbox{for} & \HOLinline{\HOLConst{Const}\;\HOLStringLitDG{=}\;(\HOLConst{Fun}\;\HOLFreeVar{ty}\;(\HOLConst{Fun}\;\HOLFreeVar{ty}\;\HOLConst{Bool}))}\\
  \HOLinline{\HOLFreeVar{s}\;\HOLSymConst{===}\;\HOLFreeVar{t}} & \mbox{for} & \HOLinline{\HOLConst{Comb}\;(\HOLConst{Comb}\;(\HOLConst{Equal}\;(\HOLConst{typeof}\;\HOLFreeVar{s}))\;\HOLFreeVar{s})\;\HOLFreeVar{t}}
\end{array}
\]

\noindent A term is \HOLinline{\HOLConst{welltyped}} if, by the following rules, it has a type:

\begin{mathpar}
\infer{\HOLinline{(\HOLConst{Var}\;\HOLFreeVar{n}\;\HOLFreeVar{ty})\;\HOLConst{has_type}\;\HOLFreeVar{ty}}}{}
\and
\infer{\HOLinline{(\HOLConst{Const}\;\HOLFreeVar{n}\;\HOLFreeVar{ty})\;\HOLConst{has_type}\;\HOLFreeVar{ty}}}{}
\and
  \infer{\HOLinline{(\HOLConst{Comb}\;\HOLFreeVar{s}\;\HOLFreeVar{t})\;\HOLConst{has_type}\;\HOLFreeVar{rty}}}{\HOLinline{\HOLFreeVar{s}\;\HOLConst{has_type}\;(\HOLConst{Fun}\;\HOLFreeVar{dty}\;\HOLFreeVar{rty})}&\HOLinline{\HOLFreeVar{t}\;\HOLConst{has_type}\;\HOLFreeVar{dty}}}
\and
  \infer{\HOLinline{(\HOLConst{Abs}\;(\HOLConst{Var}\;\HOLFreeVar{n}\;\HOLFreeVar{dty})\;\HOLFreeVar{t})\;\HOLConst{has_type}\;(\HOLConst{Fun}\;\HOLFreeVar{dty}\;\HOLFreeVar{rty})}}{\HOLinline{\HOLFreeVar{t}\;\HOLConst{has_type}\;\HOLFreeVar{rty}}}
\end{mathpar}

If a term \HOLinline{\HOLFreeVar{tm}} is \HOLinline{\HOLConst{welltyped}}, it has a unique type denoted \HOLinline{\HOLConst{typeof}\;\HOLFreeVar{tm}}.
A \emph{type substitution}, ranged over by \HOLinline{\HOLFreeVar{\ensuremath{\Theta}}}, is a mapping from type variables to types.
We write \HOLinline{\HOLFreeVar{\ensuremath{\Theta}}\;\HOLFreeVar{ty}} and \HOLinline{\HOLFreeVar{\ensuremath{\Theta}}\;\HOLFreeVar{tm}} for the result of applying \HOLinline{\HOLFreeVar{\ensuremath{\Theta}}} to all type variables of
a type \HOLinline{\HOLFreeVar{ty}} or term \HOLinline{\HOLFreeVar{tm}}, respectively; the latter case may involve $\alpha$-conversion to avoid variable capture.
In the formalisation, type substitutions are sometimes encoded as functions and sometimes as
lists of pairs. The presentation will abstract away from this distinction.
A type is \emph{ground} if it has no type variables. A substitution \HOLinline{\HOLFreeVar{\ensuremath{\Theta}}} is ground if,
for all \HOLinline{\HOLFreeVar{ty}}, \HOLinline{\HOLFreeVar{\ensuremath{\Theta}}\;\HOLFreeVar{ty}} is ground.
We say that \HOLinline{\ensuremath{\HOLFreeVar{ty}\sb{\mathrm{1}}}} is an \emph{instance} of \HOLinline{\ensuremath{\HOLFreeVar{ty}\sb{\mathrm{2}}}} (written \HOLinline{\ensuremath{\HOLFreeVar{ty}\sb{\mathrm{2}}}\;\HOLSymConst{\ensuremath{\geq}}\;\ensuremath{\HOLFreeVar{ty}\sb{\mathrm{1}}}}) if there exists a \HOLinline{\HOLFreeVar{\ensuremath{\Theta}}} such that \HOLinline{\ensuremath{\HOLFreeVar{ty}\sb{\mathrm{1}}}\;\HOLSymConst{=}\;\HOLFreeVar{\ensuremath{\Theta}}\;\ensuremath{\HOLFreeVar{ty}\sb{\mathrm{2}}}}.

We write \HOLinline{\HOLFreeVar{x}\HOLSymConst{\HOLTokenSupBullet{}}} for the list of outermost non-built-in types that occur in a type or term~\HOLinline{\HOLFreeVar{x}}.
For example $(\mathsf{map}:(\alpha\to\mathsf{Bool})\to\alpha\,\mathsf{list}\to\mathsf{Bool}\,\mathsf{list})\HOLTokenSupBullet$ returns the list containing $\alpha$ and~$\alpha\,\mathsf{list}$.
Following Kun{\v{c}}ar and Popescu, we will often be interested in these because the
model construction gives built-in types special treatment:
\begin{holthmenv}
\HOLConst{Bool}\HOLSymConst{\HOLTokenSupBullet{}}\;\HOLTokenDefEquality{}\;[]\\
(\HOLConst{Fun}\;\HOLFreeVar{dom}\;\HOLFreeVar{rng})\HOLSymConst{\HOLTokenSupBullet{}}\;\HOLTokenDefEquality{}\;\HOLFreeVar{dom}\HOLSymConst{\HOLTokenSupBullet{}}\;\HOLSymConst{++}\;\HOLFreeVar{rng}\HOLSymConst{\HOLTokenSupBullet{}}\\
\HOLFreeVar{ty}\HOLSymConst{\HOLTokenSupBullet{}}\;\HOLTokenDefEquality{}\;[\HOLFreeVar{ty}]\quad\mbox{otherwise}\\[0.5em]
(\HOLConst{Var}\;\ensuremath{\HOLFreeVar{v}\sb{\mathrm{0}}}\;\HOLFreeVar{ty})\HOLSymConst{\HOLTokenSupBullet{}}\;\HOLTokenDefEquality{}\;\HOLFreeVar{ty}\HOLSymConst{\HOLTokenSupBullet{}}\\
(\HOLConst{Const}\;\ensuremath{\HOLFreeVar{v}\sb{\mathrm{1}}}\;\HOLFreeVar{ty})\HOLSymConst{\HOLTokenSupBullet{}}\;\HOLTokenDefEquality{}\;\HOLFreeVar{ty}\HOLSymConst{\HOLTokenSupBullet{}}\\
(\HOLConst{Comb}\;\HOLFreeVar{a}\;\HOLFreeVar{b})\HOLSymConst{\HOLTokenSupBullet{}}\;\HOLTokenDefEquality{}\;\HOLFreeVar{a}\HOLSymConst{\HOLTokenSupBullet{}}\;\HOLSymConst{++}\;\HOLFreeVar{b}\HOLSymConst{\HOLTokenSupBullet{}}\\
(\HOLConst{Abs}\;\HOLFreeVar{a}\;\HOLFreeVar{b})\HOLSymConst{\HOLTokenSupBullet{}}\;\HOLTokenDefEquality{}\;\HOLFreeVar{a}\HOLSymConst{\HOLTokenSupBullet{}}\;\HOLSymConst{++}\;\HOLFreeVar{b}\HOLSymConst{\HOLTokenSupBullet{}}
\end{holthmenv}
Note that \HOLinline{\HOLConst{Bool}} and \HOLinline{\HOLTokenLambda{}\HOLBoundVar{s}\;\HOLBoundVar{t}.\;\HOLConst{Fun}\;\HOLBoundVar{s}\;\HOLBoundVar{t}} are abbreviations.
\subsection{Inference system}

A \emph{signature} is a mapping from constant names to their most general type,
and from type constructor names to their arity.
A \emph{theory} is a pair \HOLinline{(\HOLFreeVar{s}\HOLSymConst{,}\HOLFreeVar{a})} of a signature \HOLinline{\HOLFreeVar{s}} and a set of axioms
$\HOLinline{\HOLFreeVar{a}}:\HOLinline{\HOLTyOp{term}\;\ensuremath{\Rightarrow}\;\HOLTyOp{bool}}$. These components are accessed through the
self-explanatory selectors \HOLConst{axsof}, \HOLConst{tmsof}, \HOLConst{tysof} and
\HOLConst{sigof}.

We write the sequent \HOLinline{(\HOLFreeVar{thy}\HOLSymConst{,}\HOLFreeVar{h})\;\HOLSymConst{\ensuremath{\vdash}}\;\HOLFreeVar{p}} to denote that
$\HOLinline{\HOLFreeVar{p}}:\HOLinline{\HOLTyOp{term}}$ can be inferred in the theory \HOLinline{\HOLFreeVar{thy}} from the hypotheses $\HOLinline{\HOLFreeVar{h}}:\HOLinline{\HOLTyOp{term}\;\HOLTyOp{list}}$.
We define $\vdash$ as an inductive relation comprised of the standard inference rules
of higher-order logic, plus whatever axioms are present in the theory.
We show a few rules to give the flavour:
\begin{mathpar}

\infer[\HOLRuleName{ASSUME}]{\HOLinline{(\HOLFreeVar{thy}\HOLSymConst{,}[\HOLFreeVar{p}])\;\HOLSymConst{\ensuremath{\vdash}}\;\HOLFreeVar{p}}}{\HOLinline{\HOLConst{theory_ok}\;\HOLFreeVar{thy}}&\HOLinline{\HOLFreeVar{p}\;\HOLConst{has_type}\;\HOLConst{Bool}}&\HOLinline{\HOLConst{term_ok}\;(\HOLConst{sigof}\;\HOLFreeVar{thy})
\;\;\HOLFreeVar{p}}}
\and
\infer[\HOLRuleName{AXIOM}]{\HOLinline{(\HOLFreeVar{thy}\HOLSymConst{,}[])\;\HOLSymConst{\ensuremath{\vdash}}\;\HOLFreeVar{c}}}{\HOLinline{\HOLConst{theory_ok}\;\HOLFreeVar{thy}}&\HOLinline{\HOLFreeVar{c}\;\HOLSymConst{\HOLTokenIn{}}\;\HOLConst{axsof}\;\HOLFreeVar{thy}}}
\and
\infer[\HOLRuleName{MK_COMB}]{\HOLinline{(\HOLFreeVar{thy}\HOLSymConst{,}\;\ensuremath{\HOLFreeVar{h}\sb{\mathrm{1}}}\;\ensuremath{\cup}\;\ensuremath{\HOLFreeVar{h}\sb{\mathrm{2}}})\;\HOLSymConst{\ensuremath{\vdash}}\;\HOLConst{Comb}\;\ensuremath{\HOLFreeVar{l}\sb{\mathrm{1}}}\;\ensuremath{\HOLFreeVar{l}\sb{\mathrm{2}}}\;\HOLSymConst{===}\;\HOLConst{Comb}\;\ensuremath{\HOLFreeVar{r}\sb{\mathrm{1}}}\;\ensuremath{\HOLFreeVar{r}\sb{\mathrm{2}}}}}{\HOLinline{(\HOLFreeVar{thy}\HOLSymConst{,}\ensuremath{\HOLFreeVar{h}\sb{\mathrm{1}}})\;\HOLSymConst{\ensuremath{\vdash}}
\ensuremath{\HOLFreeVar{l}\sb{\mathrm{1}}}\;\HOLSymConst{===}\;\ensuremath{\HOLFreeVar{r}\sb{\mathrm{1}}}}&\HOLinline{(\HOLFreeVar{thy}\HOLSymConst{,}\ensuremath{\HOLFreeVar{h}\sb{\mathrm{2}}})\;\HOLSymConst{\ensuremath{\vdash}}\;\ensuremath{\HOLFreeVar{l}\sb{\mathrm{2}}}\;\HOLSymConst{===}\;\ensuremath{\HOLFreeVar{r}\sb{\mathrm{2}}}}&\HOLinline{\HOLConst{welltyped}\;(\HOLConst{Comb}\;\ensuremath{\HOLFreeVar{l}\sb{\mathrm{1}}}\;\ensuremath{\HOLFreeVar{l}\sb{\mathrm{2}}})}}

\end{mathpar}
Here \HOLinline{\HOLConst{term_ok}} and \HOLinline{\HOLConst{theory_ok}} are syntactic well-formedness conditions. For terms, all types and constants occurring in them must be part of the signature. For theories, all axioms must be \HOLinline{\HOLConst{term_ok}} and have type \HOLinline{\HOLConst{Bool}}, and the signature must contain the built-in types: functions, booleans and equality.

\subsection{Theory extension}\label{sec:thyext}

A theory is constructed incrementally by applying a sequence of \emph{updates}:
\begin{holthmenv}
\HOLTyOp{update}\;=\\
\;\;\;\;\HOLConst{ConstSpec}\;\HOLTyOp{bool}\;((\HOLTyOp{string}\;\HOLTokenProd{}\;\HOLTyOp{term})\;\HOLTyOp{list})\;\HOLTyOp{term}\\
\;\;\HOLTokenBar{}\;\HOLConst{TypeDefn}\;\HOLTyOp{string}\;\HOLTyOp{term}\;\HOLTyOp{string}\;\HOLTyOp{string}\\
\;\;\HOLTokenBar{}\;\HOLConst{NewType}\;\HOLTyOp{string}\;\HOLTyOp{num}\\
\;\;\HOLTokenBar{}\;\HOLConst{NewConst}\;\HOLTyOp{string}\;\HOLTyOp{type}\\
\;\;\HOLTokenBar{}\;\HOLConst{NewAxiom}\;\HOLTyOp{term}
\end{holthmenv}
\HOLinline{\HOLConst{NewType}} and \HOLinline{\HOLConst{NewConst}} introduce new type constructors and constant names into the theory signature without introducing any axioms about them.
\HOLinline{\HOLConst{TypeDefn}\;\HOLFreeVar{name}\;\HOLFreeVar{pred}\;\HOLFreeVar{abs}\;\HOLFreeVar{rep}} defines a new type \HOLinline{\HOLFreeVar{name}} by carving out from an existing type the subset that satisfies the predicate \HOLinline{\HOLFreeVar{pred}}. It also introduces abstraction and representation functions for the new type, along with two axioms: $\HOLConst{abs}(\HOLConst{rep}\,a) = a$ and $\HOLConst{pred}\,r = (\HOLConst{rep}(\HOLConst{abs}\,r) = r)$.

Constant specification is the first place where we depart significantly from Kumar et al. by allowing ad-hoc overloading; we will defer its discussion until we have introduced a few preliminaries.

A \emph{context} is a list of updates. The relation \HOLinline{\HOLSymConst{updates}} defines what constitutes a valid update of a context. We show the rules for introducing unchecked axioms (\HOLConst{NewAxiom}) and type definitions:

\begin{mathpar}
\infer{\HOLinline{\HOLConst{NewAxiom}\;\HOLFreeVar{prop}\;\HOLConst{updates}\;\HOLFreeVar{ctxt}}}{\HOLinline{\HOLFreeVar{prop}\;\HOLConst{has_type}\;\HOLConst{Bool}}&\HOLinline{\HOLConst{term_ok}
\;(\HOLConst{sigof}\;\HOLFreeVar{ctxt})\;\HOLFreeVar{prop}}}

\and


\inferrule*
           {\HOLinline{(\HOLConst{thyof}\;\HOLFreeVar{ctxt}\HOLSymConst{,}[])\;\HOLSymConst{\ensuremath{\vdash}}\;
\HOLConst{Comb}\;\HOLFreeVar{pred}\;\HOLFreeVar{witness}}\\\HOLinline{\HOLConst{closed}\;\HOLFreeVar{pred}}\\\HOLinline{\HOLFreeVar{name}\;\HOLSymConst{\HOLTokenNotIn{}}\;\HOLConst{domain}\;(\HOLConst{tysof}\;\HOLFreeVar{ctxt})}\\\HOLinline{\HOLFreeVar{abs}\;\HOLSymConst{\HOLTokenNotIn{}}\;
\HOLConst{domain}\;(\HOLConst{tmsof}\;\HOLFreeVar{ctxt})}\\\HOLinline{\HOLFreeVar{rep}\;\HOLSymConst{\HOLTokenNotIn{}}\;\HOLConst{domain}\;(\HOLConst{tmsof}\;\HOLFreeVar{ctxt})}\\\HOLinline{\HOLFreeVar{abs}\;\HOLSymConst{\HOLTokenNotEqual{}}\;\HOLFreeVar{rep}}}
           {\HOLinline{\HOLConst{TypeDefn}\;\HOLFreeVar{name}\;\HOLFreeVar{pred}\;\HOLFreeVar{abs}\;\HOLFreeVar{rep}\;\HOLConst{updates}\;\HOLFreeVar{ctxt}}}

\end{mathpar}

In words, an axiom can be introduced if it is an ok boolean term. A type definition can be introduced if it can be proved that the set carved out by \HOLinline{\HOLFreeVar{pred}} is inhabited by a \HOLinline{\HOLFreeVar{witness}},
if \HOLinline{\HOLFreeVar{pred}} has no free term variables (\HOLConst{closed}), and the names of the type and its abstraction and representation function are fresh.

We say that \HOLinline{\ensuremath{\HOLFreeVar{ctxt}\sb{\mathrm{1}}}\;\HOLConst{extends}\;\ensuremath{\HOLFreeVar{ctxt}\sb{\mathrm{2}}}} if \HOLinline{\ensuremath{\HOLFreeVar{ctxt}\sb{\mathrm{1}}}} can be obtained from
\HOLinline{\ensuremath{\HOLFreeVar{ctxt}\sb{\mathrm{2}}}} by applying a (possibly empty) sequence of valid updates.
We say that \HOLinline{\ensuremath{\HOLFreeVar{ctxt}\sb{\mathrm{1}}}} is a \HOLinline{\HOLConst{definitional_extension}} of \HOLinline{\ensuremath{\HOLFreeVar{ctxt}\sb{\mathrm{2}}}} if, additionally,
none of the updates to \HOLinline{\ensuremath{\HOLFreeVar{ctxt}\sb{\mathrm{2}}}} are \HOLinline{\HOLConst{NewAxiom}}s.
We will be particularly interested in definitional extensions of three pre-defined initial contexts:

\begin{itemize}
  \item \HOLinline{\HOLConst{init_ctxt}} introduces the built-ins---booleans, equality and functions---using \HOLinline{\HOLConst{NewType}} and \HOLinline{\HOLConst{NewConst}}. This is the bare minimum setup required for the inference system.
  \item \HOLinline{\HOLConst{finite_hol_ctxt}} extends \HOLinline{\HOLConst{init_ctxt}} with the theory of booleans, the choice constant \HOLinline{\HOLStringLitDG{@}}, the axiom
    of extensionality, and the axiom of choice (the latter two with \HOLinline{\HOLConst{NewAxiom}}).
  \item \HOLinline{\HOLConst{hol_ctxt}} extends \HOLinline{\HOLConst{finite_hol_ctxt}} with the type of individuals and the axiom of infinity.
\end{itemize}


\subsubsection{Definitional dependencies}

Obua~\cite{DBLP:conf/rta/Obua06} noticed that ad-hoc overloading is safe if definitions
do not overlap, and if unfolding of definitions terminates; we follow Kun\v{c}ar and Popescu
in formalising the former as \emph{orthogonality}, and the latter as cycle-freedom of a
dependency relation.

\paragraph{Orthogonality} A context is \emph{orthogonal} if definitions do not overlap, that is, if all constants and types have at most one definition at every type instance.
Two types $\HOLinline{\ensuremath{\HOLFreeVar{ty}\sb{\mathrm{1}}}},\HOLinline{\ensuremath{\HOLFreeVar{ty}\sb{\mathrm{2}}}}$ are orthogonal, written \HOLinline{\ensuremath{\HOLFreeVar{ty}\sb{\mathrm{1}}}\;\HOLSymConst{\ensuremath{\#}}\;\ensuremath{\HOLFreeVar{ty}\sb{\mathrm{2}}}}, if they do not have any common type instance.
\begin{holthmenv}
\ensuremath{\HOLFreeVar{ty}\sb{\mathrm{1}}}\;\HOLSymConst{\ensuremath{\#}}\;\ensuremath{\HOLFreeVar{ty}\sb{\mathrm{2}}}\;\HOLTokenDefEquality{}\;\HOLSymConst{\HOLTokenNeg{}}\HOLSymConst{\HOLTokenExists{}}\HOLBoundVar{ty}.\;\ensuremath{\HOLFreeVar{ty}\sb{\mathrm{1}}}\;\HOLSymConst{\ensuremath{\geq}}\;\HOLBoundVar{ty}\;\HOLSymConst{\HOLTokenConj{}}\;\ensuremath{\HOLFreeVar{ty}\sb{\mathrm{2}}}\;\HOLSymConst{\ensuremath{\geq}}\;\HOLBoundVar{ty}
\end{holthmenv}
Orthogonality extends to constants by looking at the types if the constant names are alike:
\begin{holthmenv}
\HOLConst{Const}\;\HOLFreeVar{c}\;\ensuremath{\HOLFreeVar{ty}\sb{\mathrm{1}}}\;\HOLSymConst{\ensuremath{\#}}\;\HOLConst{Const}\;\HOLFreeVar{d}\;\ensuremath{\HOLFreeVar{ty}\sb{\mathrm{2}}}\;\HOLTokenDefEquality{}\;\HOLFreeVar{c}\;\HOLSymConst{\HOLTokenNotEqual{}}\;\HOLFreeVar{d}\;\HOLSymConst{\HOLTokenDisj{}}\;\ensuremath{\HOLFreeVar{ty}\sb{\mathrm{1}}}\;\HOLSymConst{\ensuremath{\#}}\;\ensuremath{\HOLFreeVar{ty}\sb{\mathrm{2}}}
\end{holthmenv}
A context is orthogonal if all defined types and constants are pairwise orthogonal.

\paragraph{Cycle freedom} Every context induces a dependency relation \HOLinline{\HOLConst{\ensuremath{\rightsquigarrow}}} on \HOLinline{\HOLTyOp{type}\;\ensuremath{+}\;\HOLTyOp{term}}. The intent is that \HOLinline{\dependencypair{\HOLFreeVar{ctxt}}{\HOLFreeVar{s}}{\HOLFreeVar{t}}} means that the definition of \HOLinline{\HOLFreeVar{s}} depends directly on the definition of \HOLinline{\HOLFreeVar{t}} in the definitional theory \HOLinline{\HOLFreeVar{ctxt}}.

The \emph{substitutive closure} \HOLinline{\HOLConst{subst_clos}\;\HOLFreeVar{\ensuremath{\mathcal{R}}}} of a relation \HOLinline{\HOLFreeVar{\ensuremath{\mathcal{R}}}} relates
\HOLinline{\HOLConst{\!\!}\;\HOLFreeVar{\ensuremath{\Theta}}\;\ensuremath{\HOLFreeVar{t}\sb{\mathrm{1}}}} and \HOLinline{\HOLConst{\!\!}\;\HOLFreeVar{\ensuremath{\Theta}}\;\ensuremath{\HOLFreeVar{t}\sb{\mathrm{2}}}} whenever $\HOLinline{\ensuremath{\HOLFreeVar{t}\sb{\mathrm{1}}}}\mathrel{\mathcal R}\HOLinline{\ensuremath{\HOLFreeVar{t}\sb{\mathrm{2}}}}$
for all $\HOLinline{\HOLFreeVar{\ensuremath{\Theta}}},\HOLinline{\ensuremath{\HOLFreeVar{t}\sb{\mathrm{1}}}},\HOLinline{\ensuremath{\HOLFreeVar{t}\sb{\mathrm{2}}}}$. We say that a relation $\mathcal{R}$ is \HOLConst{terminating} if there is no sequence $x$ such that $x_i \mathrel{\mathcal{R}} x_{i+1}$ for all $i\in\mathbb{N}$. Note that if a relation is terminating, its inverse is well-founded.
Cycle freedom holds whenever the substitutive closure of the dependency relation is terminating.

In the definition of the dependency relation, we consider constants occuring within terms. The function \HOLinline{\HOLConst{\ensuremath{\cdot^\circ}}} gives the argument term's non-built-in constants.
\begin{holthmenv}
(\HOLConst{Var}\;\HOLFreeVar{x}\;\HOLFreeVar{ty})\HOLSymConst{\HOLTokenSupCirc{}}\;\HOLTokenDefEquality{}\;[]\\
(\HOLConst{Comb}\;\HOLFreeVar{a}\;\HOLFreeVar{b})\HOLSymConst{\HOLTokenSupCirc{}}\;\HOLTokenDefEquality{}\;\HOLFreeVar{a}\HOLSymConst{\HOLTokenSupCirc{}}\;\HOLSymConst{++}\;\HOLFreeVar{b}\HOLSymConst{\HOLTokenSupCirc{}}\\
(\HOLConst{Abs}\;\HOLFreeVar{\HOLTokenUnderscore{}}\;\HOLFreeVar{a})\HOLSymConst{\HOLTokenSupCirc{}}\;\HOLTokenDefEquality{}\;\HOLFreeVar{a}\HOLSymConst{\HOLTokenSupCirc{}}\\
(\HOLConst{Equal}\;\HOLFreeVar{ty})\HOLSymConst{\HOLTokenSupCirc{}}\;\HOLTokenDefEquality{}\;[]\\
(\HOLConst{Const}\;\HOLFreeVar{c}\;\HOLFreeVar{ty})\HOLSymConst{\HOLTokenSupCirc{}}\;\HOLTokenDefEquality{}\;[\HOLConst{Const}\;\HOLFreeVar{c}\;\HOLFreeVar{ty}]\quad\mbox{otherwise}
\end{holthmenv}

The dependency relation of Kun\v{c}ar and Popescu is inductively defined via two rules, where $u\equiv{}t$ means a definition with definiendum $u$ defined by a term $t$ and either $u$ is a constant (introduced through $\mathsf{ConstSpec}$) or a type (introduced by $\mathsf{TypeDefn}$). In the following cases \HOLinline{\dependencypair{\HOLFreeVar{ctxt}}{\HOLFreeVar{u}}{\HOLFreeVar{v}}} holds:
\begin{enumerate}
  \item \label{itm:one} There is a constant or type definition $u\equiv{}t$ in \HOLinline{\HOLFreeVar{ctxt}} such that \HOLinline{\HOLFreeVar{v}\;\HOLConst{\HOLTokenIn{}}\;\HOLFreeVar{t}\HOLSymConst{\HOLTokenSupBullet{}}}, or \HOLinline{\HOLFreeVar{v}\;\HOLConst{\HOLTokenIn{}}\;\HOLFreeVar{t}\HOLSymConst{\HOLTokenSupCirc{}}}; otherwise,
  \item \label{itm:two}
    $u$ is a constant-instance of type $\sigma$
    and \HOLinline{\HOLFreeVar{v}\;\HOLConst{\HOLTokenIn{}}\;\HOLFreeVar{\ensuremath{\sigma}}\HOLSymConst{\HOLTokenSupBullet{}}}
\end{enumerate}

In our mechanisation, the definition of the dependency relation has seven cases. The first five encode \ref{itm:one} and \ref{itm:two}.
Case \ref{itm:two} is very close to its mechanisation. Whenever a constant is declared, it introduces dependencies on its type's non-built-in subtypes:
\begin{mathpar}
\infer{\HOLinline{\dependencypair{\;\HOLFreeVar{ctxt}\;}{\;\HOLConst{INR}\;(\HOLConst{Const}\;\HOLFreeVar{name}\;\ensuremath{\HOLFreeVar{t}\sb{\mathrm{2}}})}{\;\HOLConst{INL}\;\ensuremath{\HOLFreeVar{t}\sb{\mathrm{1}}}\;}}}{\HOLinline{\ensuremath{\HOLFreeVar{t}\sb{\mathrm{1}}}\;\HOLConst{\HOLTokenIn{}}\;\ensuremath{\HOLFreeVar{t}\sb{\mathrm{2}}}\HOLSymConst{\HOLTokenSupBullet{}}}&\HOLinline{\HOLConst{NewConst}\;\HOLFreeVar{name}\;\ensuremath{\HOLFreeVar{t}\sb{\mathrm{2}}}\;\HOLConst{\HOLTokenIn{}}\;\HOLFreeVar{ctxt}}}
\end{mathpar}

The other two rules are absent in Kun\v{c}ar and Popescu's
dependency relation.
The first states that a type declaration \HOLinline{\HOLConst{NewType}\;\HOLFreeVar{name}\;\HOLFreeVar{arity}} (without a definition) introduces dependencies on all of the---possibly distinct---arguments to the constructor.
\begin{mathpar}
\inferrule*{
  \HOLinline{\HOLConst{NewType}\;\HOLFreeVar{name}\;\HOLFreeVar{arity}\;\HOLConst{\HOLTokenIn{}}\;\HOLFreeVar{ctxt}}\\
  \HOLinline{\HOLFreeVar{tyname}\;\HOLConst{\HOLTokenIn{}}\;\HOLFreeVar{tynames}}\\
  \mbox{\HOLinline{\HOLFreeVar{tynames}\;\HOLSymConst{=}\;\;\HOLConst{genlist}\;(\HOLTokenLambda{}\HOLBoundVar{x}.\;\HOLConst{implode}\;(\HOLConst{replicate}\;(\HOLConst{Suc}\;\HOLBoundVar{x})\;\HOLCharLit{a}))\;\HOLFreeVar{arity}}}
}{
  \HOLinline{\dependencypair{\;\HOLFreeVar{ctxt}\;}{\;\HOLConst{INL}\;(\HOLConst{Tyapp}\;\HOLFreeVar{name}\;(\HOLConst{map}\;\HOLConst{Tyvar}\;\HOLFreeVar{tynames}))}{\\
\;\HOLConst{INL}\;(\HOLConst{Tyvar}\;\HOLFreeVar{tyname})}}
}
\end{mathpar}
That is, a constructor \HOLinline{\HOLConst{Tyapp}\;\HOLFreeVar{name}\;(\HOLConst{map}\;\HOLConst{Tyvar}\;\HOLFreeVar{tynames})} depends on each of its distinctly named type variables. As we shall see in Section~\ref{sec:model}, it is crucial that this clause applie to the built-in types: since built-ins such as \HOLConst{Fun} are introduced into the initial
context with \HOLinline{\HOLConst{NewType}}, this rule gives us the ability to track dependencies that occur
as type parameters of the built-ins.

Second, a type definition \HOLinline{\HOLConst{TypeDefn}\;\HOLFreeVar{name}\;\HOLFreeVar{pred}\;\HOLFreeVar{abs}\;\HOLFreeVar{rep}} introduces the constants \HOLinline{\HOLFreeVar{abs}} and \HOLinline{\HOLFreeVar{rep}}; the final rule states that these constants depend on their types.

\subsubsection{Constant specification}

Following Candle, we admit constant specifications in the style of Arthan~\cite{DBLP:conf/itp/Arthan14}, to whom we refer for motivation, and extend this mechanism to admit ad-hoc overloading.
In a constant specification \HOLinline{\HOLConst{ConstSpec}\;\HOLFreeVar{ov}\;\HOLFreeVar{eqs}\;\HOLFreeVar{prop}}, \HOLinline{\HOLFreeVar{ov}} is a flag indicating whether the
introduced specifications are overloads. Each element $(c_i,t_i)$ of the list \HOLinline{\HOLFreeVar{eqs}} denotes the defining equation
  \HOLinline{\HOLConst{Const}\;c_i\;(\HOLConst{typeof}\;t_i)\;\HOLSymConst{===}\;t_i}
of the constant $c_i$. These equations are \emph{not} added as axioms to the context; instead, \HOLinline{\HOLConst{ConstSpec}} allows more abstraction by adding the user's choice of any \HOLinline{\HOLFreeVar{prop}} as an axiom, provided the user can prove that \HOLinline{\HOLFreeVar{prop}} follows from the defining equations.

Formally, the rule for introducing new specifications is:

\begin{mathpar}

\inferrule*{\HOLinline{(\HOLConst{thyof}\;\HOLFreeVar{ctxt}\HOLSymConst{,}
\;\HOLConst{map}\;(\HOLTokenLambda{}(\HOLBoundVar{s}\HOLSymConst{,}\HOLBoundVar{t}).\;\HOLConst{Var}\;\HOLBoundVar{s}\;(\HOLConst{typeof}\;\HOLBoundVar{t})\;\HOLSymConst{===}\;\HOLBoundVar{t})\;\HOLFreeVar{eqs})\;\HOLSymConst{\ensuremath{\vdash}}\;\HOLFreeVar{prop}}\\
\HOLinline{
\HOLSymConst{\HOLTokenForall{}}(\HOLBoundVar{c}\HOLSymConst{,}\HOLBoundVar{t})\;\HOLSymConst{\ensuremath{\in}}\;\HOLFreeVar{eqs}.\;
\HOLConst{closed}\;\HOLBoundVar{t}\;\HOLSymConst{\HOLTokenConj{}}
\;\HOLSymConst{\HOLTokenForall{}}\HOLBoundVar{v}.\;
\HOLBoundVar{v}\;\HOLSymConst{\ensuremath{\in}}\;\HOLConst{tvars}\;\HOLBoundVar{t}
\;\HOLSymConst{\HOLTokenImp{}}\;
\HOLBoundVar{v}\;\HOLSymConst{\ensuremath{\in}}\;\HOLConst{tyvars}\;(\HOLConst{typeof}\;\HOLBoundVar{t})}\\
\HOLinline{\HOLSymConst{\HOLTokenForall{}}\HOLBoundVar{x}\;\HOLBoundVar{ty}.
\;\HOLConst{Var}\;\HOLBoundVar{x}\;\HOLBoundVar{ty}\;\HOLSymConst{\ensuremath{\in}}\;\HOLConst{fv}\;\HOLFreeVar{prop}\;\HOLSymConst{\HOLTokenImp{}}
\;(\HOLBoundVar{x}\HOLSymConst{,}\HOLBoundVar{ty})\;\HOLSymConst{\ensuremath{\in}}\;\HOLConst{map}\;(\HOLTokenLambda{}(\HOLBoundVar{s}\HOLSymConst{,}\HOLBoundVar{t}).\;(\HOLBoundVar{s}\HOLSymConst{,}\HOLConst{typeof}\;\HOLBoundVar{t}))\;\HOLFreeVar{eqs}}\\\HOLinline{\HOLConst{constspec_ok}
\;\HOLFreeVar{ov}\;\HOLFreeVar{eqs}\;\HOLFreeVar{prop}\;\HOLFreeVar{ctxt}}}
{\HOLinline{\HOLConst{ConstSpec}\;\HOLFreeVar{ov}\;\HOLFreeVar{eqs}\;\HOLFreeVar{prop}\;\HOLConst{updates}\;\HOLFreeVar{ctxt}}}

\end{mathpar}

Three conditions must hold regardless of whether we are introducing overloads or not. First, the \HOLinline{\HOLFreeVar{prop}} must follow from the equations, as described above. Second, for each defining equation the RHS must have no free term variables, and may not mention type variables except those on the LHS. Third, the \HOLinline{\HOLFreeVar{prop}} must have no free term variables except those on the LHS of \HOLinline{\HOLFreeVar{eqs}}.%
\footnote{After the \HOLinline{\HOLConst{ConstSpec}} is complete, these variables will be promoted to constants.}
The requisites that differ depending on whether we are introducing overloads or not are gathered in \HOLinline{\HOLConst{constspec_ok}}:
\begin{holthmenv}
\HOLConst{constspec_ok}\;\HOLFreeVar{ov}\;\HOLFreeVar{eqs}\;\HOLFreeVar{prop}\;\HOLFreeVar{ctxt}\;\HOLTokenDefEquality{}\\
\;\;\HOLKeyword{if}\;\HOLFreeVar{ov}\;\HOLKeyword{then}\\
\;\;\;\;\HOLConst{terminating}\;(\HOLConst{subst_clos}\;(\HOLConst{\ensuremath{\rightsquigarrow}}\sb{\HOLConstSub{ConstSpec}\;\HOLFreeVar{ov}\;\HOLFreeVar{eqs}\;\HOLFreeVar{prop}\HOLSymConst{::}\HOLFreeVar{ctxt}}))\;\HOLSymConst{\HOLTokenConj{}}\\
\;\;\;\;\HOLConst{orth_ctxt}\;(\HOLConst{ConstSpec}\;\HOLFreeVar{ov}\;\HOLFreeVar{eqs}\;\HOLFreeVar{prop}\HOLSymConst{::}\HOLFreeVar{ctxt})\;\HOLSymConst{\HOLTokenConj{}}\\
\;\;\;\;\HOLSymConst{\HOLTokenForall{}}\HOLBoundVar{name}\;\HOLBoundVar{trm}.\\
\;\;\;\;\;\;\;\;(\HOLBoundVar{name}\HOLSymConst{,}\HOLBoundVar{trm})\HOLConst{\HOLTokenIn{}}\;\HOLFreeVar{eqs}\;\HOLSymConst{\HOLTokenImp{}}\\
\;\;\;\;\;\;\;\;\;\;\HOLSymConst{\HOLTokenExists{}}\ensuremath{\HOLBoundVar{ty}\sp{\prime}}.\\
\;\;\;\;\;\;\;\;\;\;\;\;\;\;\HOLConst{NewConst}\;\HOLBoundVar{name}\;\ensuremath{\HOLBoundVar{ty}\sp{\prime}}\;\HOLConst{\HOLTokenIn{}}\;\HOLFreeVar{ctxt}\;\HOLSymConst{\HOLTokenConj{}}\;\ensuremath{\HOLBoundVar{ty}\sp{\prime}}\;\HOLSymConst{\ensuremath{\geq}}\;\HOLConst{typeof}\;\HOLBoundVar{trm}\;\HOLSymConst{\HOLTokenConj{}}\\
\;\;\;\;\;\;\;\;\;\;\;\;\;\;\HOLConst{alookup}\;(\HOLConst{const_list}\;\HOLFreeVar{ctxt})\;\HOLBoundVar{name}\;\HOLSymConst{=}\;\HOLConst{Some}\;\ensuremath{\HOLBoundVar{ty}\sp{\prime}}\;\HOLSymConst{\HOLTokenConj{}}\;\HOLSymConst{\HOLTokenNeg{}}\HOLConst{is_reserved_name}\;\HOLBoundVar{name}\\
\;\;\HOLKeyword{else}\;\HOLConst{all_distinct}\;(\HOLConst{map}\;\HOLConst{fst}\;\HOLFreeVar{eqs})\;\HOLSymConst{\HOLTokenConj{}}\;\HOLSymConst{\HOLTokenForall{}}\HOLBoundVar{s}.\;\HOLBoundVar{s}\;\HOLConst{\HOLTokenIn{}}\;\HOLConst{map}\;\HOLConst{fst}\;\HOLFreeVar{eqs}\;\HOLSymConst{\HOLTokenImp{}}\;\HOLBoundVar{s}\;\HOLSymConst{\HOLTokenNotIn{}}\;\HOLConst{domain}\;(\HOLConst{tmsof}\;\HOLFreeVar{ctxt})
\end{holthmenv}
If we are not introducing overloads (the \HOLConst{else} branch), all specified constants must be fresh in the context and have pairwise distinct names.
Otherwise, if we are introducing overloads: the substitutive closure of its dependency relation must be terminating, the resulting context must be orthogonal, and for each new defining equation,
the constant name must not be reserved, and a more general type instance of all constants must
already be present in the signature via a \HOLinline{\HOLConst{NewConst}}.
The reserved names are for equality \HOLinline{\HOLStringLitDG{=}} and Hilbert choice \HOLinline{\HOLStringLitDG{@}}. Since these are axiomatised,
we cannot allow overloaded definitions of them because such definitions might contradict the axioms.
Note that we only need the \HOLinline{\HOLConst{terminating}} check when we introduce overloads, because we prove that no other theory extensions can create cycles.


This separate treatment of overloaded and non-overloaded constant specification
could perhaps be unified, but the separation is motivated by practical concerns.
First, we want to give users the freedom to specify through $\HOLinline{\HOLFreeVar{ov}} = \HOLinline{\HOLConst{F}}$ that certain polymorphic constants
cannot be overloaded.
For example, the axiom of infinity is expressed in terms of two non-builtin polymorphic constants,
namely existential and universal quantification.
Overloads on these would change the meaning of the axiom of infinity at certain type
instances, which means all bets are off in terms of consistency.
Second, it simplifies definitions and proofs if we know whether an update
extends the signature or not independently of which context it updates.

\section{Semantics}
\label{sec:semantics}

In this section, we introduce the semantic domain in which we model HOL, and define what it
means for a sequent to be true in such a model.

\subsection{Semantic domain}
\label{sec:semdomain}

Following Pitts~\cite{pitts1993} and subsequent work, we use a universe of sets satisfying Zermelo's axioms as our semantic domain. By G\"{o}del's second incompleteness theorem we cannot construct such a semantic domain within HOL. Harrison obtains such a construction by augmenting HOL with axioms that grant access to large cardinals~\cite{DBLP:conf/cade/Harrison06}.
We follow an alternative approach due to Arthan~\cite{spc002} (to whom we refer for more details),
where rather than making an explicit construction of a semantic domain, we define what properties it must have. That is, definitions and
theorems that make reference to set theory are parameterised on a type variable \HOLinline{\ensuremath{\mathcal{U}}}
denoting the universe of sets, and a term variable $\HOLinline{\HOLFreeVar{mem}}:\HOLinline{\ensuremath{\mathcal{U}}\;\ensuremath{\Rightarrow}\;\ensuremath{\mathcal{U}}\;\ensuremath{\Rightarrow}\;\HOLTyOp{bool}}$ denoting its membership relation.
The predicate \HOLinline{\HOLConst{is_set_theory}\;\HOLFreeVar{mem}} asserts that \HOLinline{\HOLFreeVar{mem}} obeys the Zermelo axioms except choice and infinity, and the predicate \HOLinline{\HOLConst{is_infinite}\;\HOLFreeVar{mem}\;\HOLFreeVar{indset}} asserts that \HOLinline{\HOLFreeVar{indset}} is an element of \HOLinline{\ensuremath{\mathcal{U}}} with infinitely many members.
We do not need to assume anything about the structure of \HOLinline{\HOLFreeVar{indset}}.
It is convenient to separate out the axiom of infinity because without it, \HOLinline{\HOLConst{is_set_theory}} can be witnessed in HOL4. We do not need the axiom of choice in our set theory because the meta-language (HOL4) has it already.

The advantage of this approach is that definitions and proofs are independent of any particular universe construction, that it is clear which theorems require such a construction, and that there is no need to pollute the global HOL environment with axiomatic extensions.
Note that this approach has implications for the trusted computing base of our mechanisation.
We trust that HOL4 faithfully implements classical higher-order
logic with choice, infinity and extensionality as described by Pitts~\cite{pitts1993}.
Additionally, we trust that the Zermelo axioms as expressed by \HOLinline{\HOLConst{is_infinite}} and
\HOLinline{\HOLConst{is_set_theory}} are consistent.
If desired, it is possible to introduce sufficient axioms into HOL4 so that \HOLinline{\HOLConst{is_infinite}} can be witnessed at a later point;
we do not do so in this paper, but see Kumar et al.~\cite{DBLP:conf/itp/KumarAMO14} for a discussion.

\paragraph{Notation}

We use the following notation for certain standard constructions in set theory.
\HOLinline{\HOLFreeVar{x}\;\HOLSymConst{\HOLTokenIn{}:}\;\HOLFreeVar{y}} abbreviates \HOLinline{\HOLFreeVar{mem}\;\HOLFreeVar{x}\;\HOLFreeVar{y}}. \HOLinline{\HOLConst{One}} is a singleton set, and \HOLinline{\HOLConst{Boolset}} is a set
with exactly two distinct elements, called \HOLinline{\HOLConst{True}} and \HOLinline{\HOLConst{False}}.
$\HOLConst{Boolean}:\HOLinline{\HOLTyOp{bool}\;\ensuremath{\Rightarrow}\;\ensuremath{\mathcal{U}}}$ maps HOL4 booleans into \HOLinline{\HOLConst{Boolset}} in the obvious way.

\HOLinline{\HOLFreeVar{x}\;\HOLConst{suchthat}\;\HOLFreeVar{P}}, where $\HOLinline{\HOLFreeVar{x}}:\HOLinline{\ensuremath{\mathcal{U}}}$ and
$\HOLinline{\HOLFreeVar{P}}:\HOLinline{\ensuremath{\mathcal{U}}\;\ensuremath{\Rightarrow}\;\HOLTyOp{bool}}$,
is the set of all elements of \HOLinline{\HOLFreeVar{x}} that satisfy \HOLinline{\HOLFreeVar{P}}.
\HOLinline{\HOLConst{Funspace}\;\HOLFreeVar{x}\;\HOLFreeVar{y}}, where $\HOLinline{\HOLFreeVar{x}},\HOLinline{\HOLFreeVar{y}}:\HOLinline{\ensuremath{\mathcal{U}}}$, is the set of all function graphs with domain \HOLinline{\HOLFreeVar{x}} and codomain \HOLinline{\HOLFreeVar{y}}.
\HOLinline{\HOLConst{Abstract}\;\HOLFreeVar{x}\;\HOLFreeVar{y}\;\HOLFreeVar{f}}, where $\HOLinline{\HOLFreeVar{f}}:\HOLinline{\ensuremath{\mathcal{U}}\;\ensuremath{\Rightarrow}\;\ensuremath{\mathcal{U}}}$, is the subset of \HOLinline{\HOLFreeVar{x}\;\HOLSymConst{\ensuremath{\times}}\;\HOLFreeVar{y}} such that
all its elememts have the form \HOLinline{(\HOLFreeVar{a}\HOLSymConst{,}\HOLFreeVar{f}\;\HOLFreeVar{a})} for some \HOLinline{\HOLFreeVar{a}}. If for all such \HOLinline{\HOLFreeVar{a}} it
holds that \HOLinline{\HOLFreeVar{f}\;\HOLFreeVar{a}\;\HOLSymConst{\HOLTokenIn{}:}\;\HOLFreeVar{y}}, this construction is the function graph of \HOLinline{\HOLFreeVar{f}} and an element
of \HOLinline{\HOLConst{Funspace}\;\HOLFreeVar{x}\;\HOLFreeVar{y}}.
We write \HOLinline{\HOLFreeVar{b}\;\HOLConst{'}\;\HOLFreeVar{a}} for function application: if \HOLinline{\HOLFreeVar{b}} is a function graph \HOLinline{\HOLConst{Abstract}\;\HOLFreeVar{x}\;\HOLFreeVar{y}\;\HOLFreeVar{f}} and \HOLinline{\HOLFreeVar{a}\;\HOLSymConst{\HOLTokenIn{}:}\;\HOLFreeVar{x}}, then \HOLinline{\HOLFreeVar{b}\;\HOLConst{'}\;\HOLFreeVar{a}\;\HOLSymConst{=}\;\HOLFreeVar{f}\;\HOLFreeVar{a}}.

\subsection{Fragments and fragment-localised semantics}
\label{sec:fragsemantics}

 Kun{\v{c}}ar and Popescu's style of semantics departs from standard HOL semantics in two main
 ways. First, it does not interpret type variables. Hence, for the most part, we only need to consider interpretations of ground terms and ground types. Later, we will see that type variables are interpreted in a point-wise manner for every possible ground instantiation.

Second, the semantics is \emph{fragment-localised}: since overloading may introduce dependencies that do not follow the order in which terms and types were introduced, it is sometimes necessary to give semantics to terms on the RHS of defining equations before all types and constants in the signature have an interpretation, or in other words, to give semantics to fragments of the signature. A \emph{signature fragment} \HOLinline{(\HOLFreeVar{tys}\HOLSymConst{,}\HOLFreeVar{consts})} of a signature \HOLinline{\HOLFreeVar{sig}} is comprised of subsets of the (non-built-in) ground types (\HOLinline{\HOLFreeVar{tys}}) and constants (\HOLinline{\HOLFreeVar{consts}}) of \HOLinline{\HOLFreeVar{sig}}. We require that the fragment is self-contained: for every \HOLinline{(\HOLFreeVar{c}\HOLSymConst{,}\HOLFreeVar{ty})\;\HOLSymConst{\HOLTokenIn{}}\;\HOLFreeVar{consts}}, \HOLinline{\HOLFreeVar{ty}} must be constructed using only \HOLinline{\HOLFreeVar{tys}} and the built-in type constructors \HOLConst{Fun} and \HOLinline{\HOLConst{Bool}}:
\begin{holthmenv}
  \HOLConst{is_sig_fragment}\;\HOLFreeVar{sig}\;(\HOLFreeVar{tys}\HOLSymConst{,}\HOLFreeVar{consts})\;\HOLTokenDefEquality{}\\
\;\;\HOLFreeVar{tys}\;\HOLSymConst{\HOLTokenSubset{}}\;\HOLConst{ground_types}\;\HOLFreeVar{sig}\;\HOLSymConst{\HOLTokenConj{}}\;\HOLFreeVar{tys}\;\HOLSymConst{\HOLTokenSubset{}}\;\HOLConst{nonbuiltin_types}\;\HOLSymConst{\HOLTokenConj{}}\\
\;\;\HOLFreeVar{consts}\;\HOLSymConst{\HOLTokenSubset{}}\;\HOLConst{ground_consts}\;\HOLFreeVar{sig}\;\HOLSymConst{\HOLTokenConj{}}\;\HOLFreeVar{consts}\;\HOLSymConst{\HOLTokenSubset{}}\;\HOLConst{nonbuiltin_constinsts}\;\HOLSymConst{\HOLTokenConj{}}\\
\;\;\HOLSymConst{\HOLTokenForall{}}\HOLBoundVar{s}\;\HOLBoundVar{c}.\;(\HOLBoundVar{s}\HOLSymConst{,}\HOLBoundVar{c})\;\HOLSymConst{\HOLTokenIn{}}\;\HOLFreeVar{consts}\;\HOLSymConst{\HOLTokenImp{}}\;\HOLBoundVar{c}\;\HOLSymConst{\HOLTokenIn{}}\;\HOLConst{builtin_closure}\;\HOLFreeVar{tys}
\end{holthmenv}
The \emph{total fragment} of a signature is the fragment comprised of all its non-built-in types and constants:
\begin{holthmenv}
  \HOLConst{total_fragment}\;\HOLFreeVar{sig}\;\HOLTokenDefEquality{}\\
\;\;(\HOLConst{ground_types}\;\HOLFreeVar{sig}\;\HOLSymConst{\HOLTokenInter{}}\;\HOLConst{nonbuiltin_types}\HOLSymConst{,}\HOLConst{ground_consts}\;\HOLFreeVar{sig}\;\HOLSymConst{\HOLTokenInter{}}\;\HOLConst{nonbuiltin_constinsts})
\end{holthmenv}
%
%
We let \HOLinline{\HOLFreeVar{\ensuremath{\delta}}} range over type interpretations \HOLinline{\HOLTyOp{type}\;\ensuremath{\Rightarrow}\;\ensuremath{\mathcal{U}}}, and
\HOLinline{\HOLFreeVar{\ensuremath{\gamma}}} range over constant interpretations \HOLinline{\HOLTyOp{string}\;\HOLTokenProd{}\;\HOLTyOp{type}\;\ensuremath{\Rightarrow}\;\ensuremath{\mathcal{U}}}.
The built-in types and constants have a standard, immutable interpretation; hence, it is convenient
to consider the extension \HOLinline{\HOLConst{ext}} of an interpretation \HOLinline{\HOLFreeVar{\ensuremath{\delta}}} to the built-ins
(written \HOLinline{\HOLConst{ext}\;\HOLFreeVar{\ensuremath{\delta}}}), and likewise for constant interpretations:
\begin{holthmenv}
\HOLConst{ext}\;\HOLFreeVar{\ensuremath{\delta}}\;\HOLConst{Bool}\;\HOLTokenDefEquality{}\;\HOLConst{Boolset}\\
\HOLConst{ext}\;\HOLFreeVar{\ensuremath{\delta}}\;(\HOLConst{Fun}\;\ensuremath{\HOLFreeVar{ty}\sb{\mathrm{1}}}\;\ensuremath{\HOLFreeVar{ty}\sb{\mathrm{2}}})\;\HOLTokenDefEquality{}\;\HOLConst{Funspace}\;(\HOLConst{ext}\;\HOLFreeVar{\ensuremath{\delta}}\;\ensuremath{\HOLFreeVar{ty}\sb{\mathrm{1}}})\;(\HOLConst{ext}\;\HOLFreeVar{\ensuremath{\delta}}\;\ensuremath{\HOLFreeVar{ty}\sb{\mathrm{2}}})\\
\HOLConst{ext}\;\HOLFreeVar{\ensuremath{\delta}}\;\HOLFreeVar{ty}\;\HOLTokenDefEquality{}\;\HOLFreeVar{\ensuremath{\delta}}\;\HOLFreeVar{ty}\quad\mbox{ otherwise}\\[0.5em]
\HOLConst{ext}\;\HOLFreeVar{\ensuremath{\delta}}\;\HOLFreeVar{\ensuremath{\gamma}}\;(\HOLStringLitDG{=}\HOLSymConst{,}\HOLConst{Fun}\;\HOLFreeVar{ty}\;(\HOLConst{Fun}\;\HOLFreeVar{ty}\;\HOLConst{Bool}))\;\HOLTokenDefEquality{}\\
\;\;\HOLConst{Abstract}\;(\HOLFreeVar{\ensuremath{\delta}}\;\HOLFreeVar{ty})\;(\HOLConst{Funspace}\;(\HOLFreeVar{\ensuremath{\delta}}\;\HOLFreeVar{ty})\;\HOLConst{Boolset})\;(\HOLTokenLambda{}\HOLBoundVar{x}.\;\HOLConst{Abstract}\;(\HOLFreeVar{\ensuremath{\delta}}\;\HOLFreeVar{ty})\;\HOLConst{Boolset}\;(\HOLTokenLambda{}\HOLBoundVar{y}.\;\HOLConst{Boolean}\;(\HOLBoundVar{x}\;\HOLSymConst{=}\;\HOLBoundVar{y})))\\
\HOLConst{ext}\;\HOLFreeVar{\ensuremath{\delta}}\;\HOLFreeVar{\ensuremath{\gamma}}\;\HOLFreeVar{ty}\;\HOLTokenDefEquality{}\;\HOLFreeVar{\ensuremath{\gamma}}\;\HOLFreeVar{ty}\quad\mbox{ otherwise}
\end{holthmenv}
We say that \HOLinline{(\HOLFreeVar{\ensuremath{\delta}}\HOLSymConst{,}\HOLFreeVar{\ensuremath{\gamma}})} is a \emph{fragment interpretation} of a fragment \HOLinline{(\HOLFreeVar{tys}\HOLSymConst{,}\HOLFreeVar{tms})}
if \HOLinline{\HOLFreeVar{\ensuremath{\delta}}} maps its types to non-empty sets, and \HOLinline{\HOLFreeVar{\ensuremath{\gamma}}} maps its constants to elements of their types' interpretation:
\begin{holthmenv}
  \HOLConst{is_type_frag_interpretation}\;\HOLFreeVar{tys}\;\HOLFreeVar{\ensuremath{\delta}}\;\HOLTokenDefEquality{}\;\HOLSymConst{\HOLTokenForall{}}\HOLBoundVar{ty}.\;\HOLBoundVar{ty}\;\HOLSymConst{\HOLTokenIn{}}\;\HOLFreeVar{tys}\;\HOLSymConst{\HOLTokenImp{}}\;\HOLConst{inhabited}\;(\HOLFreeVar{\ensuremath{\delta}}\;\HOLBoundVar{ty})\\[0.5em]
  \HOLConst{is_frag_interpretation}\;(\HOLFreeVar{tys}\HOLSymConst{,}\HOLFreeVar{tms})\;\HOLFreeVar{\ensuremath{\delta}}\;\HOLFreeVar{\ensuremath{\gamma}}\;\HOLTokenDefEquality{}\\
\;\;\HOLConst{is_type_frag_interpretation}\;\HOLFreeVar{tys}\;\HOLFreeVar{\ensuremath{\delta}}\;\HOLSymConst{\HOLTokenConj{}}\\
\;\;\HOLSymConst{\HOLTokenForall{}}(\HOLBoundVar{c}\HOLSymConst{,}\HOLBoundVar{ty}).\;(\HOLBoundVar{c}\HOLSymConst{,}\HOLBoundVar{ty})\;\HOLSymConst{\HOLTokenIn{}}\;\HOLFreeVar{tms}\;\HOLSymConst{\HOLTokenImp{}}\;\HOLFreeVar{\ensuremath{\gamma}}\;(\HOLBoundVar{c}\HOLSymConst{,}\HOLBoundVar{ty})\;\HOLSymConst{\HOLTokenIn{}:}\;\HOLConst{ext}\;\HOLFreeVar{\ensuremath{\delta}}\;\HOLBoundVar{ty}  
\end{holthmenv}
A \emph{fragment valuation}, ranged over by \HOLinline{\HOLFreeVar{v}}, assigns meaning to all variables whose
types are contained in the fragment:
\begin{holthmenv}
  \HOLConst{valuates_frag}\;\HOLFreeVar{frag}\;\HOLFreeVar{\ensuremath{\delta}}\;\HOLFreeVar{v}\;\HOLFreeVar{\ensuremath{\Theta}}\;\HOLTokenDefEquality{}\\
\;\;\HOLSymConst{\HOLTokenForall{}}\HOLBoundVar{x}\;\HOLBoundVar{ty}.\;\HOLConst{\!\!}\;\HOLFreeVar{\ensuremath{\Theta}}\;\HOLBoundVar{ty}\;\HOLSymConst{\HOLTokenIn{}}\;\HOLConst{types_of_frag}\;\HOLFreeVar{frag}\;\HOLSymConst{\HOLTokenImp{}}\;\HOLFreeVar{v}\;(\HOLBoundVar{x}\HOLSymConst{,}\HOLBoundVar{ty})\;\HOLSymConst{\HOLTokenIn{}:}\;\HOLConst{ext}\;\HOLFreeVar{\ensuremath{\delta}}\;(\HOLConst{\!\!}\;\HOLFreeVar{\ensuremath{\Theta}}\;\HOLBoundVar{ty})
\end{holthmenv}

Here we depart slightly from Kun{\v{c}}ar and Popescu by considering valuations of polymorphic terms relative to a ground type substitution \HOLinline{\HOLFreeVar{\ensuremath{\Theta}}}. Note that \HOLinline{\HOLFreeVar{v}\;(\HOLFreeVar{x}\HOLSymConst{,}\HOLFreeVar{ty})} here is \emph{not} an element of \HOLinline{\HOLFreeVar{\ensuremath{\delta}}\;\HOLFreeVar{ty}}, but of \HOLinline{\HOLFreeVar{\ensuremath{\delta}}\;(\HOLFreeVar{\ensuremath{\Theta}}\;\HOLFreeVar{ty})}.
We make this adaptation to avoid a variable capture issue. HOL has Church-style atoms, in the sense that variables with the same name but distinct types are considered to be distinct variables, so e.g.~\HOLinline{\HOLConst{Var}\;\HOLStringLitDG{x}\;(\HOLConst{Tyvar}\;\HOLStringLitDG{a})} and \HOLinline{\HOLConst{Var}\;\HOLStringLitDG{x}\;\HOLConst{Bool}} may coexist in terms and theorem statements. Because the semantics of type variables are always taken relative to a ground substitution \HOLinline{\HOLFreeVar{\ensuremath{\Theta}}}, Kun{\v{c}}ar and Popescu's definition of a fragment valuation (called an \emph{$\mathcal{I}$-compatible valuation} in~\cite{kuncar2019}) would force these two distinct \HOLinline{\HOLStringLitDG{x}} variables to have the same interpretation whenever \HOLinline{\HOLFreeVar{\ensuremath{\Theta}}\;(\HOLConst{Tyvar}\;\HOLStringLitDG{a}\;)\;\HOLSymConst{=}\;\HOLConst{Bool}}.

The same issue must be considered for the term semantics, where we lift the interpretation of a fragment's
types, constants and variables to all terms that can be built from the fragment:
\begin{holthmenv}
  \HOLConst{termsem}\;\HOLFreeVar{\ensuremath{\delta}}\;\HOLFreeVar{\ensuremath{\gamma}}\;\HOLFreeVar{v}\;\HOLFreeVar{\ensuremath{\Theta}}\;(\HOLConst{Var}\;\HOLFreeVar{x}\;\HOLFreeVar{ty})\;\HOLTokenDefEquality{}\;\HOLFreeVar{v}\;(\HOLFreeVar{x}\HOLSymConst{,}\HOLFreeVar{ty})\\
  \HOLConst{termsem}\;\HOLFreeVar{\ensuremath{\delta}}\;\HOLFreeVar{\ensuremath{\gamma}}\;\HOLFreeVar{v}\;\HOLFreeVar{\ensuremath{\Theta}}\;(\HOLConst{Const}\;\HOLFreeVar{name}\;\HOLFreeVar{ty})\;\HOLTokenDefEquality{}\;\HOLFreeVar{\ensuremath{\gamma}}\;(\HOLFreeVar{name}\HOLSymConst{,}\HOLConst{\!\!}\;\HOLFreeVar{\ensuremath{\Theta}}\;\HOLFreeVar{ty})\\
  \HOLConst{termsem}\;\HOLFreeVar{\ensuremath{\delta}}\;\HOLFreeVar{\ensuremath{\gamma}}\;\HOLFreeVar{v}\;\HOLFreeVar{\ensuremath{\Theta}}\;(\HOLConst{Comb}\;\ensuremath{\HOLFreeVar{t}\sb{\mathrm{1}}}\;\ensuremath{\HOLFreeVar{t}\sb{\mathrm{2}}})\;\HOLTokenDefEquality{}\\
\;\;\HOLConst{termsem}\;\HOLFreeVar{\ensuremath{\delta}}\;\HOLFreeVar{\ensuremath{\gamma}}\;\HOLFreeVar{v}\;\HOLFreeVar{\ensuremath{\Theta}}\;\ensuremath{\HOLFreeVar{t}\sb{\mathrm{1}}}\;\HOLConst{'}\;(\HOLConst{termsem}\;\HOLFreeVar{\ensuremath{\delta}}\;\HOLFreeVar{\ensuremath{\gamma}}\;\HOLFreeVar{v}\;\HOLFreeVar{\ensuremath{\Theta}}\;\ensuremath{\HOLFreeVar{t}\sb{\mathrm{2}}})\\
  \HOLConst{termsem}\;\HOLFreeVar{\ensuremath{\delta}}\;\HOLFreeVar{\ensuremath{\gamma}}\;\HOLFreeVar{v}\;\HOLFreeVar{\ensuremath{\Theta}}\;(\HOLConst{Abs}\;(\HOLConst{Var}\;\HOLFreeVar{x}\;\HOLFreeVar{ty})\;\HOLFreeVar{b})\;\HOLTokenDefEquality{}\\
\;\;\HOLConst{Abstract}\;(\HOLFreeVar{\ensuremath{\delta}}\;(\HOLConst{\!\!}\;\HOLFreeVar{\ensuremath{\Theta}}\;\HOLFreeVar{ty}))\;(\HOLFreeVar{\ensuremath{\delta}}\;(\HOLConst{\!\!}\;\HOLFreeVar{\ensuremath{\Theta}}\;(\HOLConst{typeof}\;\HOLFreeVar{b})))\\
\;\;\;\;(\HOLTokenLambda{}\HOLBoundVar{m}.\;\HOLConst{termsem}\;\HOLFreeVar{\ensuremath{\delta}}\;\HOLFreeVar{\ensuremath{\gamma}}\;\HOLFreeVar{v}\ensuremath{\llparenthesis}(\HOLFreeVar{x}\HOLSymConst{,}\HOLFreeVar{ty})\;\mapsto\;\HOLBoundVar{m}\ensuremath{\rrparenthesis}\;\HOLFreeVar{\ensuremath{\Theta}}\;\HOLFreeVar{b})
\end{holthmenv}
Where Kun{\v{c}}ar and Popescu apply the ground type substitution \HOLinline{\HOLFreeVar{\ensuremath{\Theta}}} eagerly, i.e.~before taking the term semantics, we
instead parameterise our term semantics on \HOLinline{\HOLFreeVar{\ensuremath{\Theta}}}, applying it only at the latest possible moment.
Crucially, we do not apply \HOLinline{\HOLFreeVar{\ensuremath{\Theta}}} at all to term variables.

We are now ready to define the meaning of sequents. An interpretation \HOLinline{(\HOLFreeVar{\ensuremath{\delta}}\HOLSymConst{,}\HOLFreeVar{\ensuremath{\gamma}})} of the fragment \HOLinline{\HOLFreeVar{frag}} satisfies the sequent \HOLinline{(\HOLFreeVar{h}\HOLSymConst{,}\HOLFreeVar{c})}
relative to the ground substitution \HOLinline{\HOLFreeVar{\ensuremath{\Theta}}} if,
whenever all the hypotheses \HOLinline{\HOLFreeVar{h}} have semantics \HOLinline{\HOLConst{True}} under a
fragment valuation \HOLinline{\HOLFreeVar{v}},
then so does the conclusion \HOLinline{\HOLFreeVar{c}}:
\begin{holthmenv}
  \HOLConst{satisfies}\;\HOLFreeVar{frag}\;\HOLFreeVar{\ensuremath{\delta}}\;\HOLFreeVar{\ensuremath{\gamma}}\;\HOLFreeVar{\ensuremath{\Theta}}\;(\HOLFreeVar{h}\HOLSymConst{,}\HOLFreeVar{c})\;\HOLTokenDefEquality{}\\
\;\;\HOLSymConst{\HOLTokenForall{}}\HOLBoundVar{v}.\\
\;\;\;\;\;\;\HOLConst{valuates_frag}\;\HOLFreeVar{frag}\;\HOLFreeVar{\ensuremath{\delta}}\;\HOLBoundVar{v}\;\HOLFreeVar{\ensuremath{\Theta}}\;\HOLSymConst{\HOLTokenConj{}}\;\HOLFreeVar{c}\;\HOLSymConst{\HOLTokenIn{}}\;\HOLConst{terms_of_frag_uninst}\;\HOLFreeVar{frag}\;\HOLFreeVar{\ensuremath{\Theta}}\;\HOLSymConst{\HOLTokenConj{}}\\
\;\;\;\;\;\;\HOLConst{every}\;(\HOLTokenLambda{}\HOLBoundVar{t}.\;\HOLBoundVar{t}\;\HOLSymConst{\HOLTokenIn{}}\;\HOLConst{terms_of_frag_uninst}\;\HOLFreeVar{frag}\;\HOLFreeVar{\ensuremath{\Theta}})\;\HOLFreeVar{h}\;\HOLSymConst{\HOLTokenConj{}}\\
\;\;\;\;\;\;\HOLConst{every}\;(\HOLTokenLambda{}\HOLBoundVar{t}.\;\HOLConst{termsem}\;\HOLFreeVar{\ensuremath{\delta}}\;\HOLFreeVar{\ensuremath{\gamma}}\;\HOLBoundVar{v}\;\HOLFreeVar{\ensuremath{\Theta}}\;\HOLBoundVar{t}\;\HOLSymConst{=}\;\HOLConst{True})\;\HOLFreeVar{h}\;\HOLSymConst{\HOLTokenImp{}}\;\HOLConst{termsem}\;\HOLFreeVar{\ensuremath{\delta}}\;\HOLFreeVar{\ensuremath{\gamma}}\;\HOLBoundVar{v}\;\HOLFreeVar{\ensuremath{\Theta}}\;\HOLFreeVar{c}\;\HOLSymConst{=}\;\HOLConst{True}
\end{holthmenv}
As promised, we then give semantics to all sequents of a signature by closing \HOLinline{\HOLConst{satisfies}} under all ground type substitutions:
\begin{holthmenv}
  \HOLConst{sat}\;\HOLFreeVar{sig}\;\HOLFreeVar{\ensuremath{\delta}}\;\HOLFreeVar{\ensuremath{\gamma}}\;(\HOLFreeVar{h}\HOLSymConst{,}\HOLFreeVar{c})\;\HOLTokenDefEquality{}\\
\;\;\HOLSymConst{\HOLTokenForall{}}\HOLBoundVar{\ensuremath{\Theta}}.\\
\;\;\;\;\;\;(\HOLSymConst{\HOLTokenForall{}}\HOLBoundVar{ty}.\;\HOLConst{tyvars}\;(\HOLBoundVar{\ensuremath{\Theta}}\;\HOLBoundVar{ty})\;\HOLSymConst{=}\;[])\;\HOLSymConst{\HOLTokenConj{}}\;(\HOLSymConst{\HOLTokenForall{}}\HOLBoundVar{ty}.\;\HOLConst{type_ok}\;(\HOLConst{tysof}\;\HOLFreeVar{sig})\;(\HOLBoundVar{\ensuremath{\Theta}}\;\HOLBoundVar{ty}))\;\HOLSymConst{\HOLTokenConj{}}\\
\;\;\;\;\;\;\HOLConst{every}\;(\HOLTokenLambda{}\HOLBoundVar{tm}.\;\HOLBoundVar{tm}\;\HOLSymConst{\HOLTokenIn{}}\;\HOLConst{ground_terms_uninst}\;\HOLFreeVar{sig}\;\HOLBoundVar{\ensuremath{\Theta}})\;\HOLFreeVar{h}\;\HOLSymConst{\HOLTokenConj{}}\\
\;\;\;\;\;\;\HOLFreeVar{c}\;\HOLSymConst{\HOLTokenIn{}}\;\HOLConst{ground_terms_uninst}\;\HOLFreeVar{sig}\;\HOLBoundVar{\ensuremath{\Theta}}\;\HOLSymConst{\HOLTokenImp{}}\\
\;\;\;\;\;\;\;\;\HOLConst{satisfies}\;(\HOLConst{total_fragment}\;\HOLFreeVar{sig})\;\HOLFreeVar{\ensuremath{\delta}}\;\HOLFreeVar{\ensuremath{\gamma}}\;\HOLBoundVar{\ensuremath{\Theta}}\;(\HOLFreeVar{h}\HOLSymConst{,}\HOLFreeVar{c})
\end{holthmenv}
We say that an interpretation \HOLinline{(\HOLFreeVar{\ensuremath{\delta}}\HOLSymConst{,}\HOLFreeVar{\ensuremath{\gamma}})} \emph{models} a theory if it interprets its total fragment and satisfies all of its axioms:
\begin{holthmenv}
  \HOLConst{models}\;\HOLFreeVar{\ensuremath{\delta}}\;\HOLFreeVar{\ensuremath{\gamma}}\;\HOLFreeVar{thy}\;\HOLTokenDefEquality{}\\
\;\;\HOLConst{is_frag_interpretation}\;(\HOLConst{total_fragment}\;(\HOLConst{sigof}\;\HOLFreeVar{thy}))\;\HOLFreeVar{\ensuremath{\delta}}\;\HOLFreeVar{\ensuremath{\gamma}}\;\HOLSymConst{\HOLTokenConj{}}\\
\;\;\HOLSymConst{\HOLTokenForall{}}\HOLBoundVar{p}.\;\HOLBoundVar{p}\;\HOLSymConst{\HOLTokenIn{}}\;\HOLConst{axsof}\;\HOLFreeVar{thy}\;\HOLSymConst{\HOLTokenImp{}}\;\HOLConst{sat}\;(\HOLConst{sigof}\;\HOLFreeVar{thy})\;(\HOLConst{ext}\;\HOLFreeVar{\ensuremath{\delta}})\;(\HOLConst{ext}\;(\HOLConst{ext}\;\HOLFreeVar{\ensuremath{\delta}})\;\HOLFreeVar{\ensuremath{\gamma}})\;([]\HOLSymConst{,}\HOLBoundVar{p})
\end{holthmenv}
Finally, we define the entailment relation \HOLinline{\HOLSymConst{\ensuremath{\vDash}}}, the semantic counterpart to \HOLinline{\HOLSymConst{\ensuremath{\vdash}}}.
In words, \HOLinline{(\HOLFreeVar{thy}\HOLSymConst{,}\HOLFreeVar{h})\;\HOLSymConst{\ensuremath{\vDash}}\;\HOLFreeVar{c}} holds if all its arguments are well-formed and if \HOLinline{(\HOLFreeVar{h}\HOLSymConst{,}\HOLFreeVar{c})} is satisfied in every model of \HOLinline{\HOLFreeVar{thy}}.
\begin{holthmenv}
  (\HOLFreeVar{thy}\HOLSymConst{,}\HOLFreeVar{h})\;\HOLSymConst{\ensuremath{\vDash}}\;\HOLFreeVar{c}\;\HOLTokenDefEquality{}\\
\;\;\HOLConst{theory_ok}\;\HOLFreeVar{thy}\;\HOLSymConst{\HOLTokenConj{}}\;\HOLConst{every}\;(\HOLConst{term_ok}\;(\HOLConst{sigof}\;\HOLFreeVar{thy}))\;(\HOLFreeVar{c}\HOLSymConst{::}\HOLFreeVar{h})\;\HOLSymConst{\HOLTokenConj{}}\\
\;\;\HOLConst{every}\;(\HOLTokenLambda{}\HOLBoundVar{p}.\;\HOLBoundVar{p}\;\HOLConst{has_type}\;\HOLConst{Bool})\;(\HOLFreeVar{c}\HOLSymConst{::}\HOLFreeVar{h})\;\HOLSymConst{\HOLTokenConj{}}\;\HOLConst{hypset_ok}\;\HOLFreeVar{h}\;\HOLSymConst{\HOLTokenConj{}}\\
\;\;\HOLSymConst{\HOLTokenForall{}}\HOLBoundVar{\ensuremath{\delta}}\;\HOLBoundVar{\ensuremath{\gamma}}.\;\HOLConst{models}\;\HOLBoundVar{\ensuremath{\delta}}\;\HOLBoundVar{\ensuremath{\gamma}}\;\HOLFreeVar{thy}\;\HOLSymConst{\HOLTokenImp{}}\;\HOLConst{sat}\;(\HOLConst{sigof}\;\HOLFreeVar{thy})\;(\HOLConst{ext}\;\HOLBoundVar{\ensuremath{\delta}})\;(\HOLConst{ext}\;(\HOLConst{ext}\;\HOLBoundVar{\ensuremath{\delta}})\;\HOLBoundVar{\ensuremath{\gamma}})\;(\HOLFreeVar{h}\HOLSymConst{,}\HOLFreeVar{c})
\end{holthmenv}
\section{Soundness}
\label{sec:soundness}
Our first main result is \emph{soundness}:
\begin{holthmenv}
  \HOLTokenTurnstile{}\;\HOLConst{is_set_theory}\;\HOLFreeVar{mem}\;\HOLSymConst{\HOLTokenImp{}}\;\HOLSymConst{\HOLTokenForall{}}\HOLBoundVar{thy}\;\HOLBoundVar{h}\;\HOLBoundVar{c}.\;(\HOLBoundVar{thy}\HOLSymConst{,}\HOLBoundVar{h})\;\HOLSymConst{\ensuremath{\vdash}}\;\HOLBoundVar{c}\;\HOLSymConst{\HOLTokenImp{}}\;(\HOLBoundVar{thy}\HOLSymConst{,}\HOLBoundVar{h})\;\HOLSymConst{\ensuremath{\vDash}}\;\HOLBoundVar{c}
\end{holthmenv}
In words, the provable sequents of a theory are satisfied in all models of the theory.
The proof comprises around 700 lines of HOL4, and is a mostly straightforward
induction on the derivation of the \HOLinline{\HOLSymConst{\ensuremath{\vdash}}} judgement.
The most notable aspect of this proof is that with the satisfiability relation of Kun{\v{c}}ar and Popescu~\cite{kuncar2019}, which features eager application of ground substitutions instead of the lazy term semantics we introduce in Section~\ref{sec:semantics}, the \textsc{ABS} case of the induction does not go through.\footnote{\textsc{ABS} is a derived rule in Kun{\v{c}}ar and Popescu, but the same problem surfaces for their \textsc{EXT} rule.
The proof step that fails is the first invocation of Lemma 11(4) in the proof of Lemma 12~\cite[p.~550]{kuncar2019},
where it does not follow that the valuations $\xi$ and $\xi'$ coincide on $\theta(\Gamma)$.
 They coincide on $\Gamma$, but this does not suffice.}
The proof obligation in the \textsc{ABS} case is:
\begin{holthmenv}
\HOLConst{is_set_theory}\;\HOLFreeVar{mem}\;\HOLSymConst{\HOLTokenImp{}}\\
\;\;\HOLSymConst{\HOLTokenForall{}}\HOLBoundVar{thy}\;\HOLBoundVar{x}\;\HOLBoundVar{ty}\;\HOLBoundVar{h}\;\HOLBoundVar{l}\;\HOLBoundVar{r}.\\
\;\;\;\;\;\;
\HOLConst{Var}\;\HOLBoundVar{x}\;\HOLBoundVar{ty}\;\HOLSymConst{\ensuremath{\not\in}}\;\HOLConst{fv}\;\HOLBoundVar{h}
\;\HOLSymConst{\HOLTokenConj{}}\;\HOLConst{type_ok}\;(\HOLConst{tysof}\;\HOLBoundVar{thy})\;\HOLBoundVar{ty}\;\HOLSymConst{\HOLTokenConj{}}\;(\HOLBoundVar{thy}\HOLSymConst{,}\HOLBoundVar{h})\;\HOLSymConst{\ensuremath{\vDash}}\;\HOLBoundVar{l}\;\HOLSymConst{===}\;\HOLBoundVar{r}\;\HOLSymConst{\HOLTokenImp{}}\\
\;\;\;\;\;\;\;\;(\HOLBoundVar{thy}\HOLSymConst{,}\HOLBoundVar{h})\;\HOLSymConst{\ensuremath{\vDash}}\;\HOLConst{Abs}\;(\HOLConst{Var}\;\HOLBoundVar{x}\;\HOLBoundVar{ty})\;\HOLBoundVar{l}\;\HOLSymConst{===}\;\HOLConst{Abs}\;(\HOLConst{Var}\;\HOLBoundVar{x}\;\HOLBoundVar{ty})\;\HOLBoundVar{r}
\end{holthmenv}
The issue here is that, after unfolding $\vDash$ and \HOLinline{\HOLConst{sat}} to obtain a ground type substitution \HOLinline{\HOLFreeVar{\ensuremath{\Theta}}}, the freshness side condition no longer applies: it does not follow that \HOLinline{\HOLConst{Var}\;\HOLFreeVar{x}\;(\HOLConst{\!\!}\;\HOLFreeVar{\ensuremath{\Theta}}\;\HOLFreeVar{ty})} is fresh in \HOLinline{\HOLFreeVar{h}}. If \HOLinline{\HOLFreeVar{h}} already contains such a variable, its valuation
becomes captured by the valuation of the bound variable \HOLinline{\HOLFreeVar{x}}, rendering the induction
hypothesis inapplicable. In our lazy semantics, \HOLinline{\HOLFreeVar{\ensuremath{\Theta}}} is not applied before the valuation,
and hence the variable capture problem does not arise.

In personal communication, Popescu acknowledges this issue and proposes an alternative fix
which involves adding additional freshness side conditions to the inference rules. That
would be less invasive in terms of semantics, but has the disadvantage of burdening the
user of the inference system with more side conditions. Our solution leaves the inference
system unchanged.

\section{Model construction}
\label{sec:model}

In the previous section, we showed that all provable sequents of a theory are satisfied in all models
of the theory. This section tackles the missing puzzle piece before we can consider consistency:
to show that if the theory is constructed by definitional extension of the pre-defined initial contexts (see Section~\ref{sec:thyext}), then a model exists.

This is the most challenging part of the work, and the part where we differ most from Kumar et al.
In the absence of overloading, the model can be constructed incrementally: when theory extension cannot change the meaning of previously introduced types and constants, any model \HOLinline{(\HOLFreeVar{\ensuremath{\delta}}\HOLSymConst{,}\HOLFreeVar{\ensuremath{\gamma}})} of the old theory can be updated to create a model of the new theory, e.g.,
\HOLinline{(\HOLFreeVar{\ensuremath{\delta}}\ensuremath{\llparenthesis}\HOLFreeVar{ty}\;\mapsto\;\HOLFreeVar{x}\ensuremath{\rrparenthesis}\HOLSymConst{,}\HOLFreeVar{\ensuremath{\gamma}}\ensuremath{\llparenthesis}\HOLFreeVar{c}\;\mapsto\;\HOLFreeVar{y}\ensuremath{\rrparenthesis})} where $\HOLinline{\HOLFreeVar{x}},\HOLinline{\HOLFreeVar{y}}$ model the new types and
constants $\HOLinline{\HOLFreeVar{ty}},\HOLinline{\HOLFreeVar{c}}$.
This has the very pleasant consequence that, when augmenting a model to accommodate a theory update,
the details of how exactly the previous model was constructed can be ignored. Hence there is no
need to explicitly write down a model for the whole theory: it suffices to prove its existence by
induction on the context.

Here, we have no such luxury: overloading can change the meaning of previously defined terms, and even types if the overloaded constant is used to define the types characteristic predicate.
Moreover, since we only interpret ground terms and types, the old model can become useless even for non-overloading extensions: signature extension causes the set of ground terms and types to grow, and
the new ones will not be covered by the model construction for the previous, smaller signature.

With incremental construction not an option, we define instead a pair of mutually recursive functions,
\HOLinline{\HOLConst{type_interpretation}} and \HOLinline{\HOLConst{term_interpretation}}. Their definition is 140 lines of HOL4 script, excluding auxiliary functions, and hence not feasible to include in full here.
The main ideas of the interpretation are standard:

\begin{itemize}
  \item A constant with no definition is interpreted as an arbitrary (Hilbert-chosen) element of
        its type's interpretation.
  \item Types with no definition are interpreted as \HOLinline{\HOLConst{One}}.
  \item A constant \HOLinline{\HOLConst{Const}\;\HOLFreeVar{c}\;(\HOLConst{\!\!}\;\HOLFreeVar{\ensuremath{\Theta}}\;\HOLFreeVar{ty})} with defining equation
        \HOLinline{\HOLConst{Const}\;\HOLFreeVar{c}\;\HOLFreeVar{ty}\;\HOLSymConst{===}\;\HOLFreeVar{t}} is interpreted as the term semantics of \HOLinline{\HOLFreeVar{t}} relative to
        \HOLinline{\HOLFreeVar{\ensuremath{\Theta}}}.
  \item A type \HOLinline{\HOLFreeVar{\ensuremath{\Theta}}\;\HOLFreeVar{ty}}, whose outermost type constructor matches
    \HOLinline{\HOLConst{TypeDefn}\;\HOLFreeVar{s}\;\HOLFreeVar{pred}\;\HOLFreeVar{abs}\;\HOLFreeVar{rep}}, is interpreted as
    \HOLinline{\HOLFreeVar{x}\;\HOLConst{suchthat}\;(\HOLTokenLambda{}\HOLBoundVar{tm}.\;\HOLFreeVar{y}\;\HOLConst{'}\;\HOLBoundVar{tm}\;\HOLSymConst{=}\;\HOLConst{True})}, where \HOLinline{\HOLFreeVar{x}} is the type semantics of
    \HOLinline{\HOLConst{\!\!}\;\HOLFreeVar{\ensuremath{\Theta}}\;(\HOLConst{domain}\;(\HOLConst{typeof}\;\HOLFreeVar{pred}))} and \HOLinline{\HOLFreeVar{y}} is the term semantics of
    \HOLinline{\HOLFreeVar{pred}} relative to \HOLinline{\HOLFreeVar{\ensuremath{\Theta}}}.
  \item If a constant matches the abstraction function of a type definition,
    and if the abstract and concrete types have (respectively) interpretations
    $\HOLinline{\ensuremath{\HOLFreeVar{\ensuremath{\delta}}\sb{\mathrm{1}}}},\HOLinline{\ensuremath{\HOLFreeVar{\ensuremath{\delta}}\sb{\mathrm{2}}}}$, then its interpretation is%
    \[
      \HOLinline{\HOLConst{Abstract}\;\ensuremath{\HOLFreeVar{\ensuremath{\delta}}\sb{\mathrm{2}}}\;\ensuremath{\HOLFreeVar{\ensuremath{\delta}}\sb{\mathrm{1}}}\\
\;\;(\HOLTokenLambda{}\HOLBoundVar{v}.\;\HOLKeyword{if}\;\HOLBoundVar{v}\;\HOLSymConst{\HOLTokenIn{}:}\;\ensuremath{\HOLFreeVar{\ensuremath{\delta}}\sb{\mathrm{1}}}\;\HOLKeyword{then}\;\HOLBoundVar{v}\;\HOLKeyword{else}\;\HOLSymConst{\HOLTokenHilbert{}}\HOLBoundVar{v}.\;\HOLBoundVar{v}\;\HOLSymConst{\HOLTokenIn{}:}\;\ensuremath{\HOLFreeVar{\ensuremath{\delta}}\sb{\mathrm{1}}})}\]
    Similarly, a constant matching the representation function is interpreted by \HOLinline{\HOLConst{Abstract}\;\ensuremath{\HOLFreeVar{\ensuremath{\delta}}\sb{\mathrm{1}}}\;\ensuremath{\HOLFreeVar{\ensuremath{\delta}}\sb{\mathrm{2}}}\;\HOLConst{I}}, where \HOLinline{\HOLConst{I}} is identity.
   \item \HOLinline{\HOLStringLitDG{ind}} is interpreted as an infinite set if the axiom of infinity is present in the theory (otherwise, any inhabited set suffices), and \HOLinline{\HOLStringLitDG{@}} is interpreted as a choice function if the axiom of choice is present.
\end{itemize}

The definition is bogged down by the bookkeeping necessary to realise the above: at every step,
the context must be scanned for matching definitions, and ground substitutions witnessing
any such match must be constructed and juggled.
The most tedious bookkeeping involves the termination argument.
First, notice that because (non-definitional) contexts may have cyclic definitions,
this model construction does not terminate in general.
We circumvent this with a trick, similar to the definition of HOL's \HOLinline{\HOLConst{WHILE}} combinator:
recursive calls are guarded by an \HOLinline{\HOLKeyword{if}} statement checking if
the dependency relation is terminating, so that the recursion is
short-circuited precisely when it is not well-founded:
\begin{holthmenv}
\HOLConst{type_interpretation}\;\HOLFreeVar{ind}\;\HOLFreeVar{ctxt}\;\HOLFreeVar{ty}\;\HOLSymConst{=}\;\HOLKeyword{if}\;\HOLSymConst{\HOLTokenNeg{}}\HOLConst{terminating}\;(\HOLConst{subst_clos}\;(\HOLConst{\ensuremath{\rightsquigarrow}}\;\HOLFreeVar{ctxt}))\;\HOLKeyword{then}\;\HOLConst{One}\;\HOLKeyword{else}\;\HOLConst{\ensuremath{\dots}}
\end{holthmenv}
The second issue is that the function must be carefully written to make sure all its
recursive calls stay within the (transitive closure of the substitutive closure of the) dependency relation. E.g., whenever
we use the term semantics in the informal explanation above, we must first construct
a theory fragment from its dependents and take the term semantics wrt.~a fragment model of that
interpretation. This termination proof is quite laborious, at 495 lines of proof script.
Through our effort to prove termination, we noticed that the dependency relation as defined by
Kun\v{c}ar and Popescu~\cite{kuncar2019} is not sufficient to show that the model construction
is well-founded: their dependency relation is only defined on non-built-in types, and hence
fails to capture dependencies in type arguments guarded by function arrows.
This seems to have gone unnoticed because in the precise spot where our proof needs this
extended dependency relation,
their proof relies on an innocent-looking but
faulty lemma stating that type instantiation commutes with~\HOLTokenSupBullet:%
\[\HOLinline{\HOLConst{set}\;(\HOLConst{\!\!}\;\HOLFreeVar{\ensuremath{\Theta}}\;\HOLFreeVar{ty})\HOLSymConst{\HOLTokenSupBullet{}}\;\HOLSymConst{=}\;\HOLTokenLeftbrace{}\HOLConst{\!\!}\;\HOLFreeVar{\ensuremath{\Theta}}\;\ensuremath{\HOLBoundVar{ty}\sp{\prime}}\;\HOLTokenBar{}\;\ensuremath{\HOLBoundVar{ty}\sp{\prime}}\;\HOLConst{\HOLTokenIn{}}\;\HOLFreeVar{ty}\HOLSymConst{\HOLTokenSupBullet{}}\HOLTokenRightbrace{}}\]
For a counterexample, choose \HOLinline{\HOLFreeVar{ty}\;\HOLSymConst{=}\;\HOLConst{Tyvar}\;\HOLStringLitDG{a}} and \HOLinline{\HOLFreeVar{\ensuremath{\Theta}}} such that \HOLinline{\HOLConst{\!\!}\;\HOLFreeVar{\ensuremath{\Theta}}\;(\HOLConst{Tyvar}\;\HOLStringLitDG{a}\;)\;\HOLSymConst{=}\;\HOLConst{Fun}\;\ensuremath{\HOLFreeVar{ty}\sp{\prime}}\;\ensuremath{\HOLFreeVar{ty}\sp{\prime}}} where \HOLinline{\ensuremath{\HOLFreeVar{ty}\sp{\prime}}} is a non-built-in ground type. The LHS evaluates to \HOLinline{\HOLTokenLeftbrace{}\ensuremath{\HOLFreeVar{ty}\sp{\prime}}\HOLTokenRightbrace{}}, but the RHS to  \HOLinline{\HOLTokenLeftbrace{}\HOLConst{Fun}\;\ensuremath{\HOLFreeVar{ty}\sp{\prime}}\;\ensuremath{\HOLFreeVar{ty}\sp{\prime}}\HOLTokenRightbrace{}}.

Our solution, as discussed in Section~\ref{sec:syntax}, is to extend the dependency relation
to account for dependencies on built-in types. From personal correspondence with Popescu,
we learn that he has independently discovered the same issue since publication.
Popescu proposes a similar fix, except that it keeps the dependency relation
unchanged, instead proving that taking the closure of the dependency relation under destruction of
built-in type constructors preserves well-foundedness. Our fix has minimal impact on our semantics
at the cost of kicking a can down the road: future verification of a cyclicity checker will
need to consider a larger dependency relation.

We write \HOLinline{\HOLConst{axioms_admissible}\;\HOLFreeVar{ctxt}} if the unchecked axioms of \HOLinline{\HOLFreeVar{ctxt}}
are at most those \HOLinline{\HOLConst{hol_ctxt}}.
The main result of this section is that for such contexts,
the interpretation defined above is indeed a model:
\begin{holthmenv}
  \HOLTokenTurnstile{}\;\HOLConst{is_set_theory}\;\HOLFreeVar{mem}\;\HOLSymConst{\HOLTokenConj{}}\;\HOLConst{inhabited}\;\HOLFreeVar{ind}\;\HOLSymConst{\HOLTokenImp{}}\\
\;\;\;\;\;\HOLSymConst{\HOLTokenForall{}}\HOLBoundVar{ctxt}.\\
\;\;\;\;\;\;\;\;\;\HOLBoundVar{ctxt}\;\HOLConst{extends}\;\HOLConst{init_ctxt}\;\HOLSymConst{\HOLTokenConj{}}\;\HOLConst{axioms_admissible}\;\HOLFreeVar{mem}\;\HOLFreeVar{ind}\;\HOLBoundVar{ctxt}\;\HOLSymConst{\HOLTokenImp{}}\\
\;\;\;\;\;\;\;\;\;\;\;\HOLConst{models}\;(\HOLConst{type_interpretation}\;\HOLFreeVar{ind}\;\HOLBoundVar{ctxt})\;(\HOLConst{term_interpretation}\;\HOLFreeVar{ind}\;\HOLBoundVar{ctxt})\;(\HOLConst{thyof}\;\HOLBoundVar{ctxt})
\end{holthmenv}
The proof is tedious: the script file dedicated to the model construction is around 5500 lines.
Recall that \HOLinline{\HOLConst{models}} has two conjuncts: the candidate model must be a total fragment
interpretation, and all axioms introduced by the context must be satisfied in the model.
The first conjunct is by well-founded induction on the definition of the dependency relation.
The second conjunct is by induction on \HOLinline{\ensuremath{\HOLFreeVar{ctxt}\sb{\mathrm{2}}}} for contexts of the form \HOLinline{\ensuremath{\HOLFreeVar{ctxt}\sb{\mathrm{1}}}\;\HOLSymConst{++}\;\ensuremath{\HOLFreeVar{ctxt}\sb{\mathrm{2}}}}.
The prefix \HOLinline{\ensuremath{\HOLFreeVar{ctxt}\sb{\mathrm{1}}}} must be carried through the induction because, as discussed above,
the model construction applies to the entire context only.
Most interesting are the axioms arising from \HOLinline{\HOLConst{ConstSpec}} and \HOLinline{\HOLConst{TypeDefn}}; let
\HOLinline{\HOLFreeVar{a}} denote such an axiom, and let \HOLinline{(\HOLFreeVar{\ensuremath{\delta}}\HOLSymConst{,}\HOLFreeVar{\ensuremath{\gamma}})} abbreviate the model under consideration.
The definition of \HOLinline{\HOLSymConst{updates}} yields
\HOLinline{(\HOLConst{thyof}\;\ensuremath{\HOLFreeVar{ctxt}\sb{\mathrm{2}}}\HOLSymConst{,}[])\;\HOLSymConst{\ensuremath{\vdash}}\;\HOLFreeVar{a}}, but soundness is not immediately useful because \HOLinline{(\HOLFreeVar{\ensuremath{\delta}}\HOLSymConst{,}\HOLFreeVar{\ensuremath{\gamma}})}
models \HOLinline{\ensuremath{\HOLFreeVar{ctxt}\sb{\mathrm{1}}}\;\HOLSymConst{++}\;\ensuremath{\HOLFreeVar{ctxt}\sb{\mathrm{2}}}}, not \HOLinline{\ensuremath{\HOLFreeVar{ctxt}\sb{\mathrm{2}}}}.
The solution is to notice that the model construction
is valid for subtheories of \HOLinline{\HOLConst{thyof}\;(\ensuremath{\HOLFreeVar{ctxt}\sb{\mathrm{1}}}\;\HOLSymConst{++}\;\ensuremath{\HOLFreeVar{ctxt}\sb{\mathrm{2}}})} that have the same signature but
fewer axioms.
Since $\vdash$ is closed under theory extension
we get \HOLinline{((\HOLConst{sigof}\;(\ensuremath{\HOLFreeVar{ctxt}\sb{\mathrm{1}}}\;\HOLSymConst{++}\;\ensuremath{\HOLFreeVar{ctxt}\sb{\mathrm{2}}})\HOLSymConst{,}\HOLConst{axsof}\;\ensuremath{\HOLFreeVar{ctxt}\sb{\mathrm{2}}})\HOLSymConst{,}[])\;\HOLSymConst{\ensuremath{\vdash}}\;\HOLFreeVar{a}},
and by the induction hypothesis, \HOLinline{\HOLConst{models}\;\HOLFreeVar{\ensuremath{\delta}}\;\HOLFreeVar{\ensuremath{\gamma}}\;(\HOLConst{sigof}\;(\ensuremath{\HOLFreeVar{ctxt}\sb{\mathrm{1}}}\;\HOLSymConst{++}\;\ensuremath{\HOLFreeVar{ctxt}\sb{\mathrm{2}}})\HOLSymConst{,}\HOLConst{axsof}\;\ensuremath{\HOLFreeVar{ctxt}\sb{\mathrm{2}}})};
by soundness we get that \HOLinline{\HOLFreeVar{a}} is satisfied in \HOLinline{(\HOLFreeVar{\ensuremath{\delta}}\HOLSymConst{,}\HOLFreeVar{\ensuremath{\gamma}})}.

\section{Consistency}
\label{sec:consistency}

We say that a theory is \emph{consistent} if it has a provable sequent and an unprovable sequent.\footnote{As the logic is classical, the existence of an unprovable sequent is equivalent to the usual definition of consistency: for all $\varphi$, at most one of $(thy,[])\vdash\varphi$ and $(thy,[])\vdash\neg\varphi$ are sequents.} The sequents are fixed for convenience:
\begin{holthmenv}
  \HOLConst{consistent_theory}\;\HOLFreeVar{thy}\;\HOLTokenDefEquality{}\\
\;\;(\HOLFreeVar{thy}\HOLSymConst{,}[])\;\HOLSymConst{\ensuremath{\vdash}}\;\HOLConst{Var}\;\HOLStringLitDG{x}\;\HOLConst{Bool}\;\HOLSymConst{===}\;\HOLConst{Var}\;\HOLStringLitDG{x}\;\HOLConst{Bool}\;\HOLSymConst{\HOLTokenConj{}}\\
\;\;\HOLSymConst{\HOLTokenNeg{}}((\HOLFreeVar{thy}\HOLSymConst{,}[])\;\HOLSymConst{\ensuremath{\vdash}}\;\HOLConst{Var}\;\HOLStringLitDG{x}\;\HOLConst{Bool}\;\HOLSymConst{===}\;\HOLConst{Var}\;\HOLStringLitDG{y}\;\HOLConst{Bool})
\end{holthmenv}
Our main result, the consistency of higher-order logic with ad-hoc overloading,
is a straightforward consequence of soundness and the existence of a model.
It is stated as follows:
\begin{holthmenv}
  \HOLTokenTurnstile{}\;\HOLConst{is_set_theory}\;\HOLFreeVar{mem}\;\HOLSymConst{\HOLTokenConj{}}\;\HOLConst{is_infinite}\;\HOLFreeVar{mem}\;\HOLFreeVar{ind}\;\HOLSymConst{\HOLTokenImp{}}\\
\;\;\;\;\;\HOLSymConst{\HOLTokenForall{}}\HOLBoundVar{ctxt}.\;\HOLConst{definitional_extension}\;\HOLBoundVar{ctxt}\;\HOLConst{hol_ctxt}\;\HOLSymConst{\HOLTokenImp{}}\;\HOLConst{consistent_theory}\;(\HOLConst{thyof}\;\HOLBoundVar{ctxt})
\end{holthmenv}
\section{Related Work}
\label{sec:related}

Our mechanisation extends the work of Kumar et al.~\cite{DBLP:conf/itp/KumarAMO14}, who prove soundness of higher-order logic to arrive at a---down to the machine code---verified HOL theorem prover.
Earlier in Section~\ref{sec:semantics} we describe the approach to defining a model for a definitional theory in set theory which they follow. Their consistency proof is via a reduction to
stateless HOL~\cite{wiedijk2011}, where the context is not updated as definitions are annotated with
definitions for each constant and type.
The journal version~\cite{DBLP:journals/jar/KumarAMO16} drops stateless HOL in favour of a
direct consistency proof.
We build upon much of their infrastructure;
for example, definitions and properties of type substitutions.
More recently, Abrahamsson used Candle to build a verified OpenTheory article checker~\cite{ABRAHAMSSON2020100530}.

Wenzel~\cite{DBLP:conf/tphol/Wenzel97} aims to close foundational gaps that arise in Isabelle through allowing type classes and overloading for HOL. Definitional extensions are claimed to be \emph{meta-safe}, i.e. conservative and \emph{realisable}. The latter means that new constants in provable terms of an extension can be replaced such that the resulting term is provable in a smaller theory. A restriction is that constant and type definitions cannot arbitrarily be mixed. Obua~\cite{DBLP:conf/rta/Obua06} states that (mixed) type definitions and overloading are safe if any two definitions with the same name are non-overlapping and unfolding definitions terminates. He also gives an algorithm, but misses that additional type definitions may lead to inconsistencies.

These issues are discovered and solved by Kun\v{c}ar and Popescu~\cite{DBLP:conf/cpp/Kuncar15,kuncar2019} whose ideas we utilise. They introduce a dependency relation to track types and constants that a definition depends on, and ultimately prove consistency of definitional theories with overloading by constructing a model with ground, fragment-localised semantics.
The soundness proof goes by recursion over a hull of the dependency relation, which takes type instances and the transitive closure into account. Their semantics are wrt.~ground terms, i.e., a formula $\phi$ holds in a model $\mathcal{I}$ if for all ground type substitutions $\theta$ and all valuations $\xi$ the term $[\theta(\phi)]^\mathcal{I}(\xi)$ evaluates to true in $\mathcal{I}$.
In subsequent work, Kun{\v{c}}ar and Popescu~\cite{kuncar2017} show proof-theoretic conservativity (and meta-safety) of HOL with overloading over initial HOL, which we call the initial context.

Gengelbach and Weber~\cite{gengelbach2017} generalise the model construction of Kun{\v{c}}ar and Popescu to give a model-theoretic conservativity result.
A definitional theory extension with overloading may affect the interpretation of terms in a model prior to extension compared with a model for the extended theory. They observe that a model for the extensions can be constructed keeping the (by the theory extension) unaffected parts of the smaller model. When updating only necessary interpretations, both models have some equal fragments. The problems that we found in Kun{\v{c}}ar and Popescu's work impact their reasoning too, but should be resolvable in the same way.

By constraining definitions to \emph{composable} dependencies, Kun{\v{c}}ar makes the dependency relation decidable~\cite{DBLP:conf/cpp/Kuncar15}. The dependency relation of a theory $D$ is composable if a type substitutive dependency chain from a definiendum $u$ to another definiendum $u'$ of $D$ exists only with constraints on the last step: only if from $u\rightsquigarrow_D^{\downarrow+}v$ either \HOLinline{\HOLFreeVar{v}\;\HOLSymConst{\ensuremath{\geq}}\;\ensuremath{\HOLFreeVar{u}\sp{\prime}}} or \HOLinline{\ensuremath{\HOLFreeVar{u}\sp{\prime}}\;\HOLSymConst{\ensuremath{\geq}}\;\HOLFreeVar{v}} holds.
In the paper the author gives a pen-and-paper proof for decidability and correctness to check acyclicity of a composable dependency relation of a definitional theory.

Carneiro~\cite{carneiro2019} agrees that correctness of definitional axioms is critical for the soundness of a logic. In his approach to bootstrapping trust into the Metamath Zero (MM0) proof system, definitions are conservative: The definiendum can be replaced by the definiens, hence soundness without definitions implies soundness of the system with definitions. As every expression in MM0 has a unique sort and there is no hierarchy of sorts, there is no support for overloading in MM0.
Contrary to ours, at present the soundness proof is not using the implementation but an abstraction.
The MM0 verification system consists of a verifier that checks proofs against a specification, and a compiler that compiles a high level proof into a low-level representation for the verifier.
Ultimately, MM0 aims for correctness of the verifier only, which is specified in its own language and not yet formally proven correct.

Another non-HOL proof assistant is Nuprl, which uses the language of untyped $\lambda$-calculus in a variant of Martin Löf's dependent type theory. The meta-theory of Nuprl is consistent relative to the consistency of the Coq theorem prover~\cite{anand2014}, which amounts to showing soundness of 70 proof rules where both types and terms are expressed in $\lambda$-calculus.
On this result, Rahli and Bickford~\cite{rahli2018} base their verified Nuprl implementation. Constants and types, called \emph{abstractions} in Nuprl, may syntactically occur prior to their definition, which enables implicit definitions as for example fix-points. Their semantics similarly treats sequents as true, if they are true for all possible extensions of the yet undefined syntax. That corresponds to Kun{\v{c}}ar and Popescu's approach of ground semantics. The verified implementation translates a proof from Nuprl into the embedding in Coq.

Sozeau et al. develop a correct verifier for the Coq kernel within Coq~\cite{sozeau2019}, and as in Coq's logic a verification task is a type check, they prove the implementation of a type checker correct. The type checker will return a proof that a term has a certain type. By Coq's extraction mechanism they obtain a verified executable type checker for the kernel of Coq.

In addition to proving consistency through a model-theoretic argument and soundness of the derivation rules for HOL, Kunčar and Popescu~\cite{kuncarcomprehending2017} arrive at a proof-theoretic/syntactic result for HOL relative to HOL with unfolded definitions, called HOLC. By translating type definitions as well as constant definitions and restricting the use of the type instantiation rule (a complication due to overloading), they achieve conservativity of HOLC relative to HOL with definitions. Consistency of HOL with ad-hoc overloading thus translates to consistency of HOLC, which in turn holds as HOL without definitions is consistent.

\section{Conclusion}

We have presented a mechanised proof of consistency for HOL with ad-hoc overloading
of constant definitions.
In doing so, we take an important first step towards a verified implementation
of an Isabelle/HOL kernel. Given executable HOL4 definitions, extraction of such an implementation is
almost routine thanks to the rich ecosystem around CakeML~\cite{DBLP:journals/jfp/MyreenO14,DBLP:conf/cade/HoAKMTN18,ESOP17}. Our termination criterion on contexts, however, is not
executable as expressed.
The main missing puzzle piece is to implement and verify Kun{\v{c}}ar's algorithm for checking orthogonality and termination of dependency relations~\cite{DBLP:conf/cpp/Kuncar15}.
This will be the topic of future work.

Our formalisation can be also be used as a staging ground for exploring other meta-theoretical properties
of Isabelle/HOL, such as safety and conservativity.
Another interesting idea is to extend our formalisation to account for other ways that
Isabelle/HOL's logic differs from the other HOLs,
such as axiomatic type classes, schematic variables and the meta logic/object logic distinction. Presently, our setting is essentially HOL Light~\cite{DBLP:conf/tphol/Harrison09a} with ad-hoc overloading.

In the meantime, our work significantly strengthens our assurance in the foundational argument
for why Isabelle/HOL is consistent.
It is almost cliché to motivate a verification effort, \emph{ex post facto}, by reference to the bugs
it uncovers. But for bugs in the consistency argument for a widely used proof assistant,
this motivation seems doubly strong.

\section{Acknowledgments}
\label{sect:acks}

We are grateful to Andrei Popescu for insightful technical discussions,
and to Oskar Abrahamsson, Yong Kiam Tan, Konrad Slind and Tjark Weber for their comments on drafts.
We thank the anonymous reviewers for their remarks.
We also thank C.F.~Liljewalchs stipendiestiftelse for supporting the second author during his visit to UNSW and CSIRO's Data61 in Sydney, Australia.

\label{sect:bib}
\bibliographystyle{plain}
\bibliography{paper}

\begin{thebibliography}{10}

\bibitem{ABRAHAMSSON2020100530}
Oskar Abrahamsson.
\newblock A verified proof checker for higher-order logic.
\newblock {\em Journal of Logical and Algebraic Methods in Programming}, page
  100530, 2020.

\bibitem{anand2014}
Abhishek Anand and Vincent Rahli.
\newblock Towards a {Formally} {Verified} {Proof} {Assistant}.
\newblock In Gerwin Klein and Ruben Gamboa, editors, {\em Interactive {Theorem}
  {Proving}}, Lecture {Notes} in {Computer} {Science}, pages 27--44, Cham,
  2014. Springer International Publishing.

\bibitem{spc002}
Rob Arthan.
\newblock {HOL} formalised: Semantics.
\newblock \url{http://lemma-one.com/ProofPower/specs/spc002.pdf}.

\bibitem{DBLP:conf/itp/Arthan14}
Rob Arthan.
\newblock {HOL} constant definition done right.
\newblock In {\em Interactive Theorem Proving - 5th International Conference,
  {ITP} 2014, Held as Part of the Vienna Summer of Logic, {VSL} 2014, Vienna,
  Austria, July 14-17, 2014. Proceedings}, pages 531--536. Springer, 2014.

\bibitem{carneiro2019}
Mario Carneiro.
\newblock Metamath {Zero}: {The} {Cartesian} {Theorem} {Prover}.
\newblock {\em arXiv:1910.10703 [cs, math]}, October 2019.
\newblock arXiv: 1910.10703.

\bibitem{PittsAM:newaas-jv}
Murdoch~J. Gabbay and Andrew~M. Pitts.
\newblock A new approach to abstract syntax with variable binding.
\newblock {\em Formal Aspects of Computing}, 13:341--363, 2001.

\bibitem{gengelbach2017}
Arve Gengelbach and Tjark Weber.
\newblock Model-theoretic {Conservative} {Extension} of {Definitional}
  {Theories}.
\newblock In {\em Proceedings of 12th {Workshop} on {Logical} and {Semantic}
  {Frameworks} with {Applications} ({LSFA} 2017)}, pages 4--16, Brasília,
  Brasil, September 2017. Elsevier.

\bibitem{godel}
Kurt G\"{o}del.
\newblock \"{U}ber formal unentscheidbare {S}\"{a}tze der {P}rincipia
  {M}athematica und verwandter {S}ysteme, {I}.
\newblock {\em Monatshefte f\"{u}r Mathematik und Physik}, pages 173--198,
  1931.

\bibitem{ESOP17}
Armaël Guéneau, Magnus~O. Myreen, Ramana Kumar, and Michael Norrish.
\newblock Verified characteristic formulae for {CakeML}.
\newblock In Hongseok Yang, editor, {\em European Symposium on Programming
  ({ESOP})}, volume 10201 of {\em LNCS}. Springer, 2017.

\bibitem{flyspeck}
Thomas~C. Hales, Mark Adams, Gertrud Bauer, Dat~Tat Dang, John Harrison,
  Truong~Le Hoang, Cezary Kaliszyk, Victor Magron, Sean McLaughlin, Thang~Tat
  Nguyen, Truong~Quang Nguyen, Tobias Nipkow, Steven Obua, Joseph Pleso, Jason
  Rute, Alexey Solovyev, An~Hoai~Thi Ta, Trung~Nam Tran, Diep~Thi Trieu, Josef
  Urban, Ky~Khac Vu, and Roland Zumkeller.
\newblock A formal proof of the {K}epler conjecture.
\newblock {\em Forum of Mathematics, Pi}, 5:e2, 2017.

\bibitem{DBLP:conf/cade/Harrison06}
John Harrison.
\newblock Towards self-verification of {HOL} {L}ight.
\newblock In Ulrich Furbach and Natarajan Shankar, editors, {\em Automated
  Reasoning, Third International Joint Conference, {IJCAR} 2006, Seattle, WA,
  USA, August 17-20, 2006, Proceedings}, volume 4130 of {\em Lecture Notes in
  Computer Science}, pages 177--191. Springer, 2006.

\bibitem{DBLP:conf/tphol/Harrison09a}
John Harrison.
\newblock {HOL} {L}ight: An overview.
\newblock In {\em Theorem Proving in Higher Order Logics, 22nd International
  Conference, TPHOLs 2009, Munich, Germany. Proceedings}, pages 60--66.
  Springer, 2009.

\bibitem{DBLP:conf/itp/2019}
John Harrison, John O'Leary, and Andrew Tolmach, editors.
\newblock {\em 10th International Conference on Interactive Theorem Proving,
  {ITP} 2019, September 9-12, 2019, Portland, OR, {USA}}, volume 141 of {\em
  LIPIcs}. Schloss Dagstuhl - Leibniz-Zentrum f{\"{u}}r Informatik, 2019.

\bibitem{DBLP:conf/cade/HoAKMTN18}
Son Ho, Oskar Abrahamsson, Ramana Kumar, Magnus~O. Myreen, Yong~Kiam Tan, and
  Michael Norrish.
\newblock Proof-producing synthesis of {CakeML} with {I/O} and local state from
  monadic {HOL} functions.
\newblock In Didier Galmiche, Stephan Schulz, and Roberto Sebastiani, editors,
  {\em Automated Reasoning - 9th International Joint Conference, {IJCAR} 2018,
  Held as Part of the Federated Logic Conference, FloC 2018, Oxford, UK, July
  14-17, 2018, Proceedings}, volume 10900 of {\em Lecture Notes in Computer
  Science}, pages 646--662. Springer, 2018.

\bibitem{DBLP:conf/itp/ImmlerRW19}
Fabian Immler, Jonas R{\"{a}}dle, and Makarius Wenzel.
\newblock Virtualization of {HOL4} in {I}sabelle.
\newblock In Harrison et~al. \cite{DBLP:conf/itp/2019}, pages 21:1--21:18.

\bibitem{DBLP:conf/sosp/KleinEHACDEEKNSTW09}
Gerwin Klein, Kevin Elphinstone, Gernot Heiser, June Andronick, David Cock,
  Philip Derrin, Dhammika Elkaduwe, Kai Engelhardt, Rafal Kolanski, Michael
  Norrish, Thomas Sewell, Harvey Tuch, and Simon Winwood.
\newblock {seL4}: formal verification of an {OS} kernel.
\newblock In {\em Proceedings of the 22nd {ACM} Symposium on Operating Systems
  Principles 2009, {SOSP} 2009, Big Sky, Montana, USA, October 11-14, 2009},
  pages 207--220. ACM, 2009.

\bibitem{DBLP:conf/itp/KumarAMO14}
Ramana Kumar, Rob Arthan, Magnus~O. Myreen, and Scott Owens.
\newblock {HOL} with definitions: Semantics, soundness, and a verified
  implementation.
\newblock In {\em Interactive Theorem Proving - 5th International Conference,
  {ITP} 2014, Held as Part of the Vienna Summer of Logic, {VSL} 2014, Vienna,
  Austria, July 14-17, 2014. Proceedings}, pages 308--324. Springer, 2014.

\bibitem{DBLP:journals/jar/KumarAMO16}
Ramana Kumar, Rob Arthan, Magnus~O. Myreen, and Scott Owens.
\newblock Self-formalisation of higher-order logic - semantics, soundness, and
  a verified implementation.
\newblock {\em J. Autom. Reasoning}, 56(3), 2016.

\bibitem{DBLP:conf/popl/KumarMNO14}
Ramana Kumar, Magnus~O. Myreen, Michael Norrish, and Scott Owens.
\newblock {CakeML}: a verified implementation of {ML}.
\newblock In {\em The 41st Annual {ACM} {SIGPLAN-SIGACT} Symposium on
  Principles of Programming Languages, {POPL} '14, San Diego, CA, USA, January
  20-21, 2014}, pages 179--192. ACM, 2014.

\bibitem{DBLP:conf/cpp/Kuncar15}
Ond{\v{r}}ej Kun{\v{c}}ar.
\newblock Correctness of {I}sabelle's cyclicity checker: Implementability of
  overloading in proof assistants.
\newblock In {\em Proceedings of the 2015 Conference on Certified Programs and
  Proofs, {CPP} 2015, Mumbai, India, January 15-17, 2015}, pages 85--94. ACM,
  2015.

\bibitem{kuncar2017}
Ond{\v{r}}ej Kun{\v{c}}ar and Andrei Popescu.
\newblock Safety and {Conservativity} of {Definitions} in {HOL} and
  {Isabelle}/{HOL}.
\newblock {\em Proc. ACM Program. Lang.}, 2(POPL):24:1--24:26, December 2017.

\bibitem{kuncar2019}
Ond{\v{r}}ej Kun{\v{c}}ar and Andrei Popescu.
\newblock A consistent foundation for {I}sabelle/{HOL}.
\newblock {\em Journal of Automated Reasoning}, 62(4):531--555, Apr 2019.

\bibitem{kuncarcomprehending2017}
Ondřej Kunčar and Andrei Popescu.
\newblock Comprehending {Isabelle}/{HOL}'s {Consistency}.
\newblock In Hongseok Yang, editor, {\em Programming {Languages} and {Systems}
  - 26th {European} {Symposium} on {Programming}, {ESOP} 2017, {Held} as {Part}
  of the {European} {Joint} {Conferences} on {Theory} and {Practice} of
  {Software}, {ETAPS} 2017, {Uppsala}, {Sweden}, {April} 22-29, 2017,
  {Proceedings}}, volume 10201 of {\em Lecture {Notes} in {Computer}
  {Science}}, pages 724--749. Springer, 2017.

\bibitem{DBLP:journals/jfp/MyreenO14}
Magnus Myreen and Scott Owens.
\newblock Proof-producing translation of higher-order logic into pure and
  stateful {ML}.
\newblock {\em J. Funct. Program.}, 24(2--3):284--315, 2014.

\bibitem{nipkow:isabelle}
Tobias Nipkow, Lawrence~C. Paulson, and Markus Wenzel.
\newblock {\em Isabelle/{HOL}: a Proof Assistant for Higher-Order Logic},
  volume 2283.
\newblock Springer, 2002.

\bibitem{DBLP:conf/rta/Obua06}
Steven Obua.
\newblock Checking conservativity of overloaded definitions in higher-order
  logic.
\newblock In {\em Term Rewriting and Applications, 17th International
  Conference, {RTA} 2006, Seattle, WA, USA, August 12-14, 2006, Proceedings},
  pages 212--226. Springer, 2006.

\bibitem{DBLP:conf/pldi/PfenningE88}
Frank Pfenning and Conal Elliott.
\newblock Higher-order abstract syntax.
\newblock In Richard~L. Wexelblat, editor, {\em PLDI}, pages 199--208. ACM,
  1988.

\bibitem{pitts1993}
Andrew~M. Pitts.
\newblock The {HOL} logic.
\newblock In M.J.C. Gordon and Tom Melham, editors, {\em Introduction to {HOL}:
  A Theorem-Proving Environment for Higher-Order Logic}, pages 191--232.
  Cambridge University Press, 1993.

\bibitem{rahli2018}
Vincent Rahli, Liron Cohen, and Mark Bickford.
\newblock A {Verified} {Theorem} {Prover} {Backend} {Supported} by a
  {Monotonic} {Library}.
\newblock In {\em {EPiC} {Series} in {Computing}}, volume~57, pages 564--582.
  EasyChair, October 2018.

\bibitem{sozeau2019}
Matthieu Sozeau, Simon Boulier, Yannick Forster, Nicolas Tabareau, and Théo
  Winterhalter.
\newblock Coq {Coq} correct! verification of type checking and erasure for
  {Coq}, in {Coq}.
\newblock {\em Proceedings of the ACM on Programming Languages},
  4(POPL):8:1--8:28, December 2019.

\bibitem{DBLP:conf/tphol/Wenzel97}
Markus Wenzel.
\newblock Type classes and overloading in higher-order logic.
\newblock In {\em Theorem Proving in Higher Order Logics, 10th International
  Conference, TPHOLs'97, Murray Hill, NJ, USA, August 19-22, 1997,
  Proceedings}, pages 307--322, 1997.

\bibitem{wiedijk2011}
Freek Wiedijk.
\newblock Stateless {HOL}.
\newblock {\em Electronic Proceedings in Theoretical Computer Science},
  53:47--61, March 2011.
\newblock arXiv: 1103.3322 version: 1.

\end{thebibliography}

\newpage

\section*{Appendix: Accessing the Mechanisation}

If you want to evaluate our artifacts, this appendix is a brief guide to the mechanisation.
The mechanisation is accessible from the CakeML repository:
 \[\mbox{\url{https://code.cakeml.org/tree/master/candle/overloading}}\]
The mechanisation currently only runs on a highly specific development version of HOL4. Here's
how to build and run it:

\begin{enumerate}
 \item Get PolyML version 5.7, following the links and instructions at \url{polyml.org}.
   Unfortunately, newer versions may induce HOL4 to crash randomly.
 \item Check out commit \texttt{9d20add95c75b8c69157845df5d5e74f24633371} from the HOL4 repository
   at \url{https://github.com/HOL-Theorem-Prover/HOL}, and follow the installation instructions
   in the \texttt{INSTALL} file. Make sure the \texttt{bin/} folder in the HOL directory is
   in your \texttt{\$PATH}.
 \item Check out commit \texttt{210f5c2ce0804027acbac59d87eab48cdeed1625} from
   the CakeML development repository\footnote{\url{https://code.cakeml.org}}, and
    run \texttt{Holmake} in the directory \texttt{candle/overloading/semantics/} to compile all proofs of our mechanisation and their dependencies.
 \item To browse the theories interactively, follow the instructions
   in the HOL4 documentation\footnote{\url{https://hol-theorem-prover.org/\#doc}} to
   set up HOL4 interaction with your choice
   of emacs or vim. If you only want to read the theories, any text editor with unicode support
   should be alright.
\end{enumerate}

\noindent The following theories within the \texttt{candle/} directory are relevant for this paper:

\begin{itemize}
  \item \texttt{overloading/syntax/holSyntaxScript.sml} corresponds to Section~\ref{sec:syntax}.
    From the same directory, \texttt{holSyntaxExtraScript.sml}
    contains many lemmas and auxiliary definitions about the syntax,
    \texttt{holBoolSyntaxScript.sml} defines the theory of booleans,
    and \texttt{holAxiomsSyntaxScript.sml} defines the theory extensions corresponding to
    the axioms of extensionality (eta), choice, and infinity.
  \item \texttt{set-theory/setSpecScript.sml} defines the predicates
    \HOLConst{is_set_theory} and \HOLConst{is_infinite} (Section~\ref{sec:semdomain}).
    These are due to Rob Arthan~\cite{spc002}.
  \item \texttt{overloading/semantics/holSemanticsScript.sml} contains the definitions from
    Section~\ref{sec:fragsemantics}.
  \item \texttt{overloading/semantics/holSoundnessScript.sml} contains the soundness proof
    reported in Section~\ref{sec:soundness}. The main theorem is called \texttt{proves_sound}.
  \item \texttt{overloading/semantics/holBoolScript.sml} proves that any model of the theory
    of booleans must give a standard semantics to the standard logical operators and quantifiers.
    It is used a lot in the proofs, but not reported on in the paper.
  \item \texttt{overloading/semantics/holExtensionScript.sml} is Section~\ref{sec:model}: it
    defines our model construction, proves termination, and proves that it is indeed a model.
    The theorem reported in Section~\ref{sec:model} is named \texttt{interpretation_is_model}.
  \item \texttt{overloading/semantics/holConsistencyScript.sml} corresponds to Section~\ref{sec:consistency}. This paper's main result resides here and is named \texttt{hol_consistent}.
\end{itemize}

\end{document}